\pdfoutput=1

\documentclass{amsart}

\usepackage[pagewise]{lineno}
 
\usepackage{amsmath}
\usepackage{amssymb}
\usepackage{paralist}
\usepackage{graphics}
\usepackage{epsfig}
\usepackage{graphicx}  
\usepackage{epstopdf}
\usepackage{stmaryrd}

\usepackage[colorlinks=true]{hyperref}
\hypersetup{urlcolor=blue, citecolor=red}
\usepackage{hyperref}

\usepackage{rotating}
\usepackage{adjustbox}
\usepackage{array}      
\usepackage{multicol}
\usepackage[bottom]{footmisc}
\usepackage{float}
\usepackage{cancel}
\usepackage[usenames,dvipsnames]{pstricks}
\usepackage{multirow} 
\usepackage{color}
\usepackage{setspace}
\usepackage{colortbl}

\usepackage{boiboites}
\usepackage{amsfonts}
\usepackage{cancel}
\usepackage{listings}
\usepackage{float}
\usepackage{fancyhdr}
\usepackage{enumerate}
\usepackage{lastpage}
\usepackage{tikz}
\usepackage[most,many]{tcolorbox}
\usepackage{multicol}
\usepackage{kantlipsum}
\usepackage{pgfplots}
\usepackage{longtable} 
\usepackage{anyfontsize}
\usepackage{tikzpagenodes}
\usepackage{pgfplotstable}

\pgfplotsset{compat=newest}

\def\currentvolume{X}
 \def\currentissue{X}
  \def\currentyear{200X}
   \def\currentmonth{XX}
    \def\ppages{X--XX}
     \def\DOI{10.3934/xx.xx.xx.xx}

\numberwithin{equation}{section}

\newcommand{\n}{\newline}

\def\be{\begin{equation}}
\def\ee{\end{equation}}

\def\ben{\begin{eqnarray}}
\def\een{\end{eqnarray}}

\def\ee{e^{\alpha z^*}}

\def\n{\newline}

\title[Predict treatment of multi-resistant {\it P\lowercase{seudomonas aeruginosa}} infection]{Dynamics and numerical simulations to predict empirical antibiotic treatment of multi-resistant {\it P\lowercase{seudomonas aeruginosa}} infection}
\author[]{}
				


\email{jlopez78@us.es}
\email{mariapar89@gmail.com}
\email{aalcudia@us.es}
\email{bbegines@us.es}
\email{caraball@us.es}
\email{epajuelo@us.es}
\email{pv1gipep@uco.es}

\thanks{$^*$ Corresponding authors: aalcudia@us.es, jlopez78@us.es}

\thanks{$^{**}$ Both authors contributed equally to this study}

\thanks{This work has been partially supported by Project 2004/0001203 from Ministerio de Educaci\'on y Ciencia (Spanish government), Project PGC2018-096540-B-I00 from Ministerio de Ciencia, Innovaci\'on y Universidades (Spanish government) and Project US-1254251 from Consejer\'ia de Econom\'ia y Conocimiento (Junta de Andaluc\'ia).}
	
\begin{document}

\maketitle

\medskip
\centerline{\scshape Javier L\'opez-de-la-Cruz$^{*,**}$}
\medskip
{\footnotesize
\centerline{Departamento de Ecuaciones Diferenciales y An\'alisis Num\'erico}
   \centerline{C/ Tarfia s/n, Facultad de Matem\'aticas, Universidad de Sevilla}
   \centerline{ 41012 Sevilla, Spain}
}
\medskip
\centerline{\scshape Mar\'ia P\'erez-Aranda$^{**}$}
\medskip
{\footnotesize
 \centerline{Departamento de Qu\'Imica Org\'anica y Farmac\'eutica and Departamento de Microbiolog\'ia y Parasitolog\'ia}
   \centerline{C/ Profesor Garc\'ia Gonz\'alez, s/n, Facultad de Farmacia. Universidad de Sevilla}
   \centerline{ 41012 Sevilla, Spain}
}
\medskip
\centerline{\scshape Ana Alcudia$^*$ and Bel\'en Begines}
\medskip
{\footnotesize
 \centerline{Departamento de Qu\'imica Org\'anica y Farmac\'eutica}
   \centerline{C/ Profesor Garc\'ia Gonz\'alez, s/n, Facultad de Farmacia. Universidad de Sevilla}
   \centerline{ 41012 Sevilla, Spain}
}
\medskip
\centerline{\scshape Tom\'as Caraballo}
\medskip
{\footnotesize
  \centerline{Departamento de Ecuaciones Diferenciales y An\'alisis Num\'erico}
   \centerline{C/ Tarfia s/n, Facultad de Matem\'aticas, Universidad de Sevilla}
   \centerline{ 41012 Sevilla, Spain}
}
\medskip
\centerline{\scshape Elo\'isa Pajuelo}
\medskip
{\footnotesize
 \centerline{Departamento de Microbiolog\'ia y Parasitolog\'ia}
    \centerline{C/ Profesor Garc\'ia Gonz\'alez, s/n, Facultad de Farmacia. Universidad de Sevilla}
   \centerline{ 41012 Sevilla, Spain}
}
\medskip
\centerline{\scshape and Pedro J. Ginel}
\medskip
{\footnotesize
 \centerline{Departamento de Medicina y Cirug\'ia Animal}
   \centerline{Edificio Hospital Cl\'inico Veterinario, Campus de Rabanales, Facultad de Veterinaria, Universidad de C\'ordoba}
   \centerline{14071 C\'ordoba, Spain}
}
\bigskip

\centerline{(Communicated by the associate editor name)}

\begin{abstract}
This work discloses an epidemiological mathematical model to predict an empirical treatment for dogs infected by {\it Pseudomonas aeruginosa}. This dangerous pathogen is one of the leading causes of multi-resistant infections and can be transmitted from dogs to humans. Numerical simulations and appropriated codes were developed using {\it Matlab} software to gather information concerning long-time dynamics of the susceptible, infected and recovered individuals. All data compiled from the mathematical model was used to provide an appropriated antibiotic sensitivity panel for this specific infection. In this study, several variables have been included in this model to predict which treatment  should be prescribed in emergency cases, when there is no time to perform an antibiogram or the cost of it could not be assumed. In particular, we highlight the use of this model aiming to become part of the convenient toolbox of Public Health research and decision-making in the design of the mitigation strategy of bacterial pathogens. 
\end{abstract}
		
\vskip0.2cm

\section{Introduction}

Multi-drug resistant infections, typically caused by a bacterium who is resistant to more than one antibiotic, are an emerging problem and constitute an important threat to public health (see \cite{RB8,AN6,VP16}), in fact, more than 700,000 people die per year due to this kind of infections currently. Latest 2019 World Health Organization report claims for urgent action to avert the antimicrobial resistance crisis and highlights that, {\it if no action is taken,  drug-resistant diseases could cause 10 million deaths each year by 2050 and damage the economy as catastrophic as the 2008-2009 global financial crisis}.\n

In this sense, these types of infections are a defiant therapeutic challenge in both human and veterinary medicine, due to the complex mechanisms that bacteria have developed to resist to antibiotics and their capacity to transmit this resistance both vertically (to their descendants) and horizontally (to other bacteria) (see \cite{RB8,AN6}).\n

In particular, {\it Pseudomonas aeruginosa} is one of the most frequent pathogen that acquires or develops antibiotic multi-resistance. This Gram-negative rod is very ubiquitous in the environment, can persist in water and soil with minimal nutritional presence and can tolerate several conditions of humidity and temperature (see \cite{ZM19,HC}).

\begin{figure}[H]
\begin{center}
\includegraphics[scale=0.19]{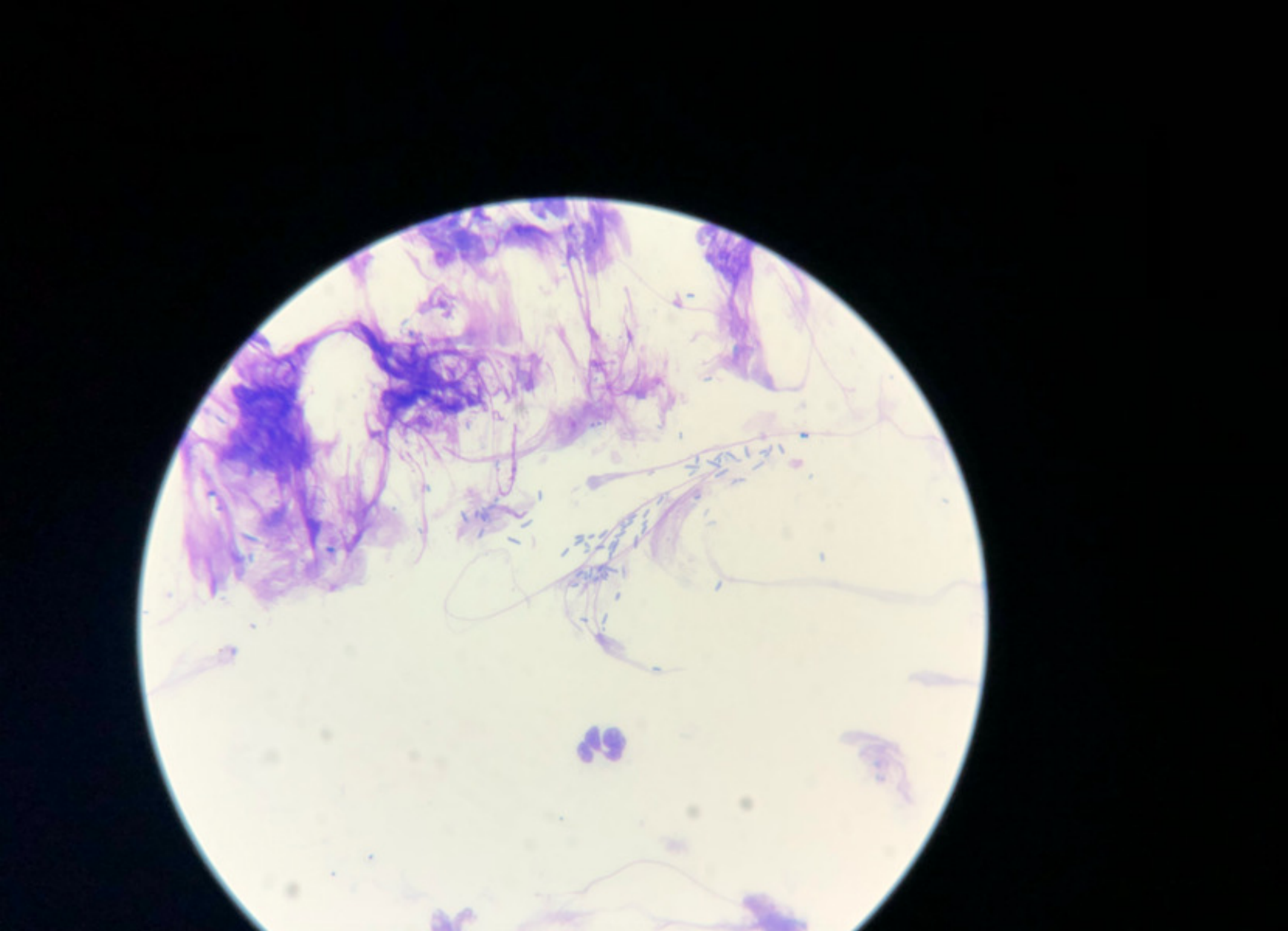}\includegraphics[scale=0.19]{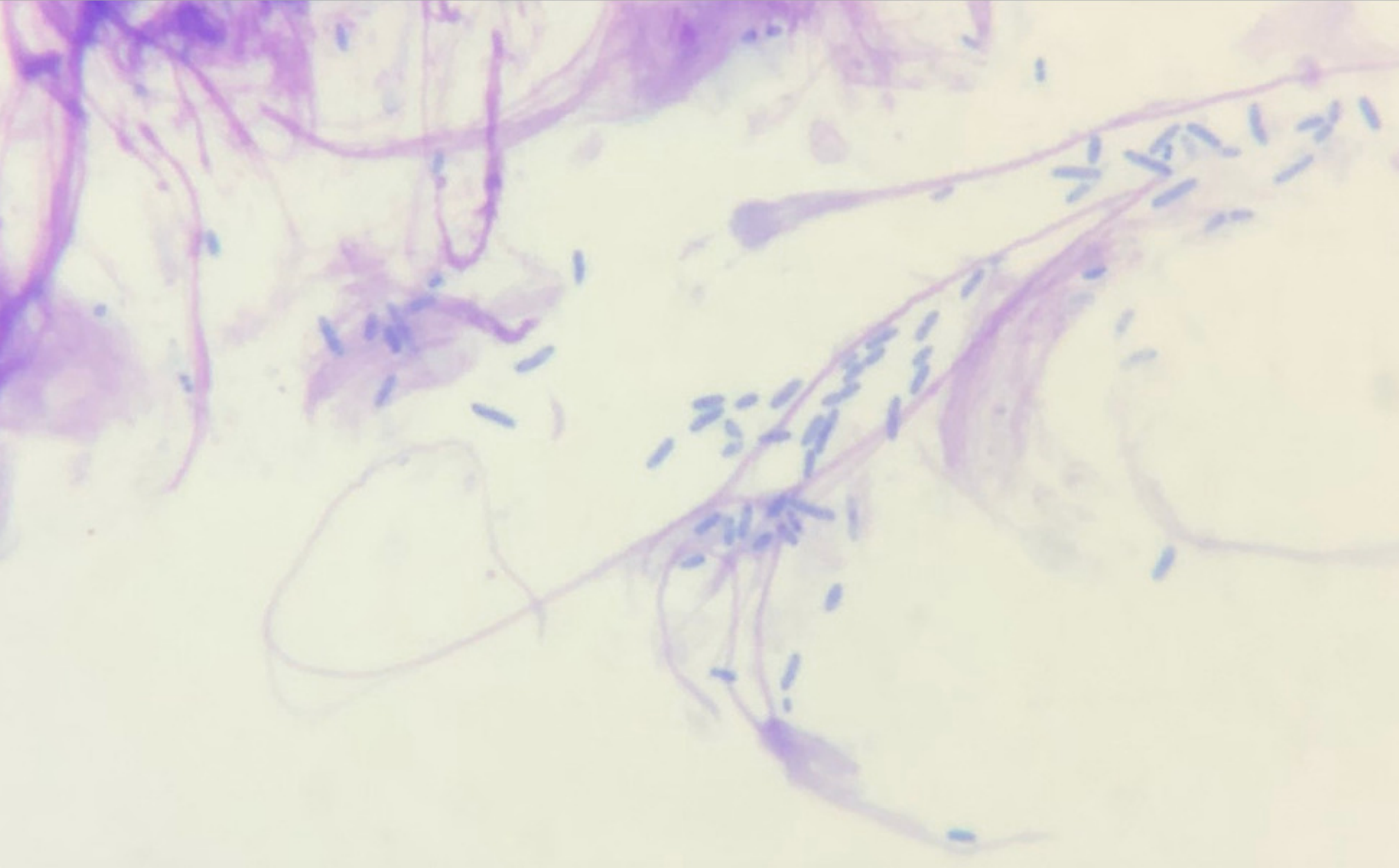}
\end{center}
\caption{Microscopic image of an exudate smear from the ear canal of a dog with otitis externa. The presence of numerous rods as well as inflammatory cells can be observed}
\label{di}
\end{figure}

As revealed, this capacity to resist extreme situations as well as the high capacity to develop multi-resistance after exposure to antimicrobials and cross-resistance between agents, makes {\it Pseudomonas aeruginosa} a leading cause of nosocomial infections, responsible for 10\% of all hospital acquired infections. In addition, these are often severe and life threatening, and very often very difficult to treat because of the limited susceptibility to antimicrobial agents (see, for instance \cite{RB8,VP16,ZM19}). In particular, in dogs this pathogen is often a cause of otitis and pyoderma. Due to the close physical contact with their owners , dogs are a potential reservoir to multi-resistant {\it Pseudomonas aeruginosa}, which could lead to zoonotic (animal to human) transmission (see \cite{HC,PR16,PY13}).\n

This work compiles 40 cases of {\it Pseudomonas aeruginosa} otitis in dogs from the south of Spain in the last five years and their susceptibility to an extended panel of antibiotics (see Table \ref{at} in Appendix). Although the best procedure to treat these infections is to perform an antibiogram study, unfortunately that takes five to seven days (see \cite{AN6,VP16,ZM19}) and not always economic resources  are available, specially in Third World Countries where sometimes the sanitary system has limited funding. Significantly, there are some critical clinical circumstances in which the treatment cannot be delayed due to the severity of the symptoms and the possible risk that may lead to an irreversible situation to the patient.

\begin{figure}[H]
\begin{center}
\includegraphics[scale=0.2]{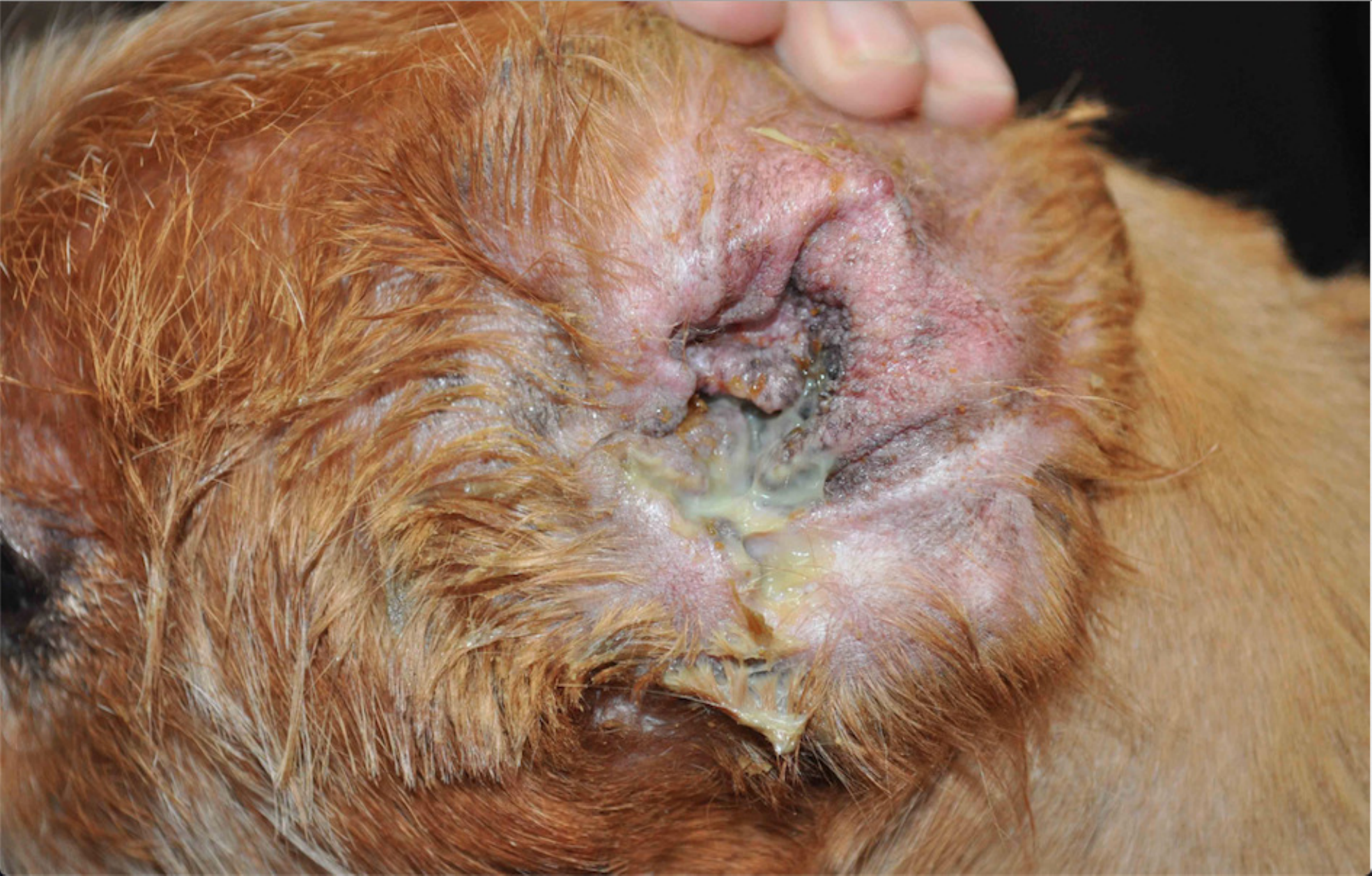}
\end{center}
\caption{Purulent otitis externa caused by Pseudomonas aeruginosa in a dog}
\label{di}
\end{figure}

In view of the facts described before, it would be crucial to have a powerful tool that could help practitioners to predict the best empirical treatment possible for this type of infection. Therefore, a professional could prescribe an antibiotic treatment as soon as the infection is detected, in case there is no time available to perform the antibiogram.\n

In this sense, the goal of the present work is to predict, using a mathematical model, the best empirical antibiotic treatment to avoid the bacterial development of resistance mechanisms.\n

The classical reference about epidemics in Mathematics is due to Kermack and McKendrick (see \cite{KM1,KM2,KM3}) who proposed a compartment model based on simple assumptions on the flow rates between different groups of members in some population. Basically, the population is divided in three groups: susceptible, the individuals who are not infected yet but susceptible to the disease; infected, who are the infected individuals; and recovered, individuals who have been infected in the past and cannot be re-infected. From now on, both groups will be denoted by $S$, $I$ and $R$, respectively.\n

Since Kermanck and McKendrick proposed the so-called SIR model, many authors have worked to develop innovations which improve the original SIR model in order to describe more complex situations \cite{cc,caraballo-book}. Among these more sophisticated models, it is worth mentioning the SIR model with relapse, or the also called SIRI model, whose main difference compared to the SIR one is based on the the fact that recovered individuals could be re-infected.\n

The SIRI model can be mathematically described with the following system of differential equations
\begin{eqnarray}
\frac{dS}{dt}&=&\Lambda(S+I+R)-\beta SI-\mu S,\label{1}
\\[1.3ex]
\frac{dI}{dt}&=&\beta SI-(\alpha+\kappa+\mu)I+\gamma R,\label{2}
\\[1.3ex]
\frac{dR}{dt}&=&\kappa I-(\mu+\gamma)R.\label{3}
\end{eqnarray}
\noindent where $S=S(t)$, $I=I(t)$ and $R=R(t)$ denote the number of susceptible, infected and recovered individuals, respectively. $\Lambda$ and $\mu$ are the birth and death rates of the populations, respectively, $\beta$ is the transmission coefficient of the disease, $\alpha$ denotes the death rate due to the disease, $\kappa$ describes the rate at which the infected individuals become recovered and $\gamma$ is the rate at which the recovered individuals become infected. We also remark that every constant is positive.\n

\begin{figure}[H]
\begin{center}
\includegraphics[width=0.7\textwidth]{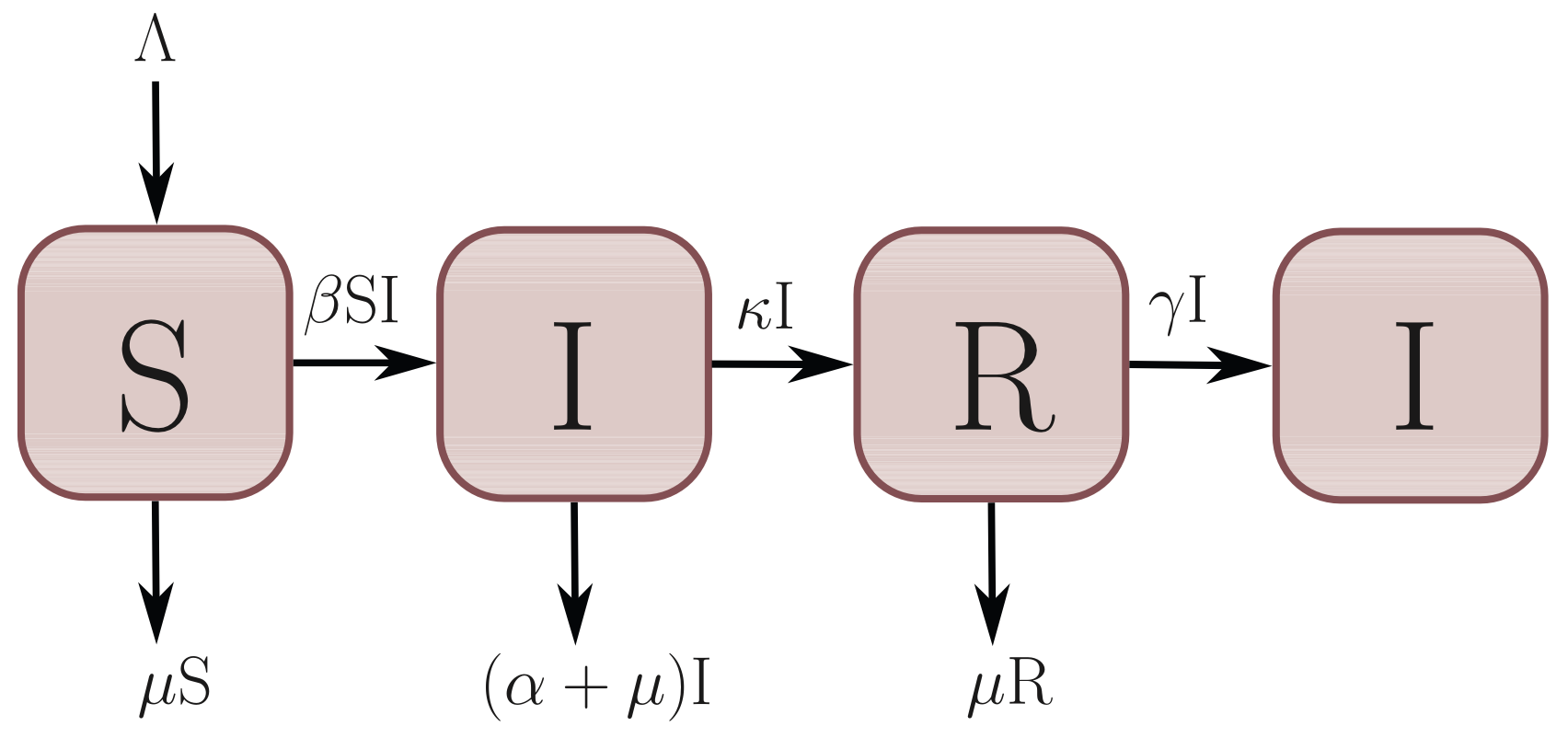}
\end{center}
\label{d}
\caption{Transfer diagram of SIRI model}
\end{figure}

The different flows between the three groups of the population can be represented as the transfer diagram in Figure \ref{d}.

To this end, we consider a population of 192 dogs of which 40 are infected and {\it Matlab} is used to make numerical simulations and understand the long-time dynamics of the different groups (susceptible, infected and recovered) of the population depending on time. In addition, we include a code in {\it Matlab} to compute the final number of susceptible, infected and recovered individuals in each case as well as the time at which it happens. Therefore, we display tables and bar diagrams compiling the information to predict the most effective antibiotics in every case.\n

The present work is organized as follows: Section \ref{material}compiles the materials and methods used in this work. Section \ref{work} shows numerical simulations of every case. All the relevant information is depicted on tables and we use bar diagrams to reach a better understanding of the dynamics. Section \ref{ad} provides a list of the most suitable antibiotics to treat infected dogs. Finally, Section \ref{finish} includes conclusions Significantly, there is a strong evidence of the utility of this type of mathematical models performed in collaboration with other disciplines, such as health sciences.
 
\section{Materials and methods}\label{material}

For this work, forty {\it Pseudomonas aeruginosa} samples were aseptically collected from dogs with purulent otitis externa and were sent to a specialized laboratory to perform their culture and antibiogram analysis. We refer the reader to Table \ref{at} in Appendix to check the whole list of antibiotics that were tested and their rates of resistance.\n

Antimicrobial susceptibility was tested using Kirby-Bauer method, also known as disk diffusion antibiotic sensitivity testing or disk diffusion test according to Clinical Laboratory Standards Institute (CLSI) guidelines (see \cite{CLSI}).

\begin{figure}[H]
\begin{center}
\includegraphics[scale=0.35]{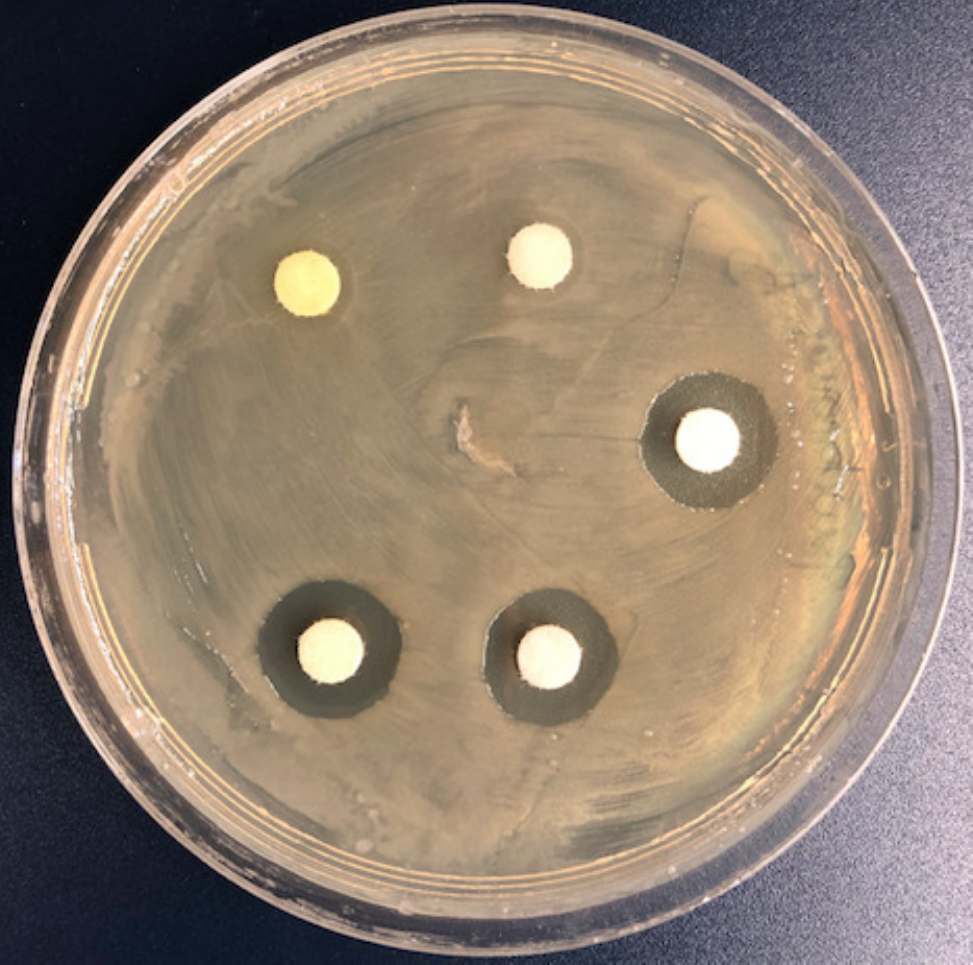}
\end{center}
\caption{Antimicrobial susceptibility assay using disk diffusions test in {\it Pseudomonas aeruginosa} culture}
\label{di}
\end{figure}

In addition, {\it Matlab} was used to make numerical simulations and program some codes to obtain useful data about the behavior of the population.

\section{Mathematical study. Numerical simulations and data}\label{work}

In this section we present the mathematical study using a SIRI model \eqref{1}-\eqref{3} predict an empirical treatment.\n

The complete panel of antibiotics tested in laboratory, as well as their corresponding resistance rate is presented in Table \ref{at} in Appendix.\n

To organize all data obtained based on the resistance rate, the following groups are depicted:

\begin{itemize}
\item {\bf Group 1-} Antibiotics 100\% resistant.
\item {\bf Group 2-} Antibiotics with resistance close to 100\%.
\item {\bf Group 3-} Antibiotics with resistance between 30\% and 50\%.
\item {\bf Group 4-} Antibiotics with resistance between 20\% and 30\%.
\item {\bf Group 5-} Antibiotics with resistance between 10\% and 20\%.
\item {\bf Group 6-} Antibiotics with resistance between 1\% and 10\%.
\item {\bf Group 7-} Antibiotics with resistance 0\% resistant.
\end{itemize}

The percentages corresponding to the groups cited before are represented in Figure \ref{di}.

\begin{figure}[H]
\begin{center}
\includegraphics[scale=0.5]{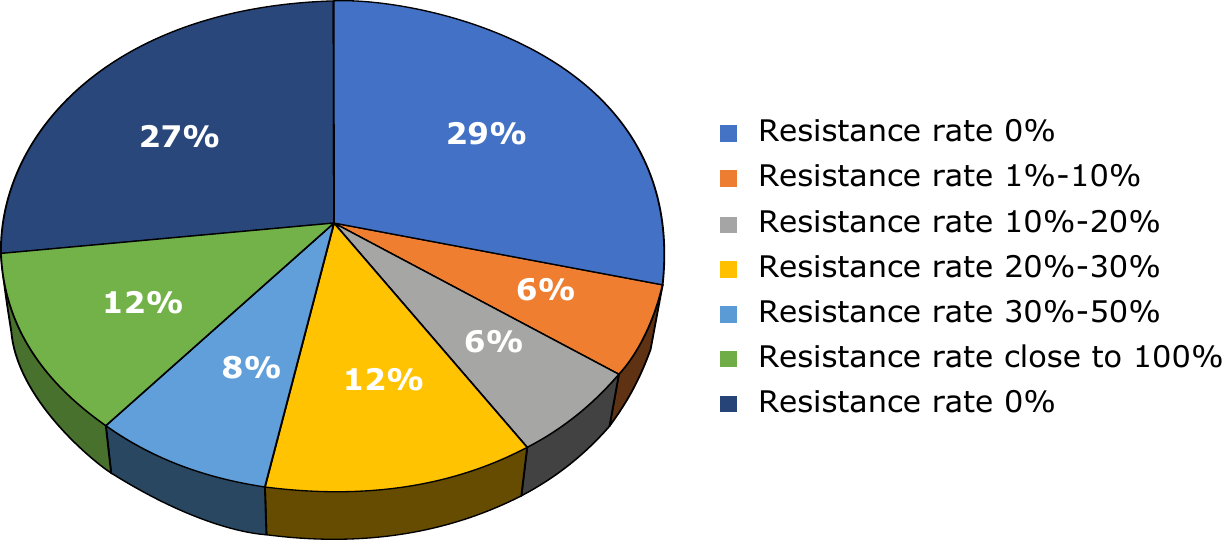}
\end{center}
\caption{Resistance rates of antimicrobials}
\label{di}
\end{figure}

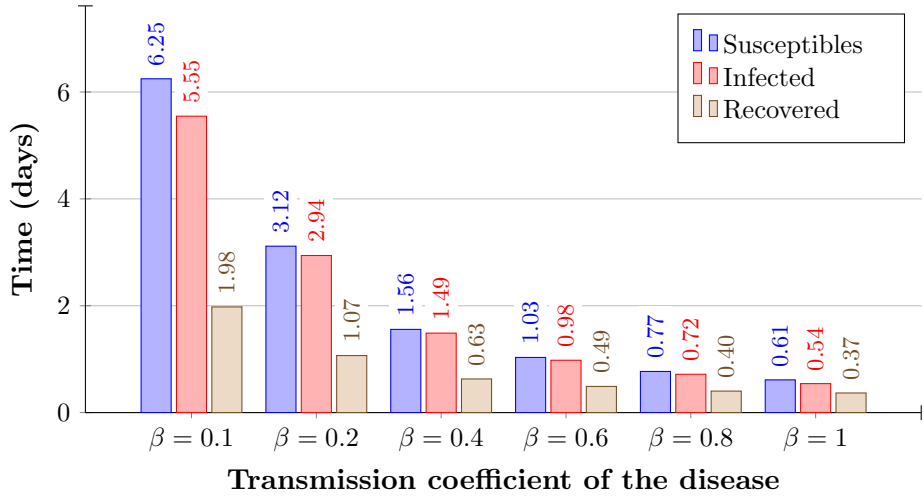
\begin{figure}[H]
\begin{center}

\begin{tikzpicture}
\pgfplotstableread{
6.2475		5.5475		1.9775
3.1150		2.9400		1.0675
1.5575		1.4875		0.6300
1.0325		0.9800		0.4900
0.7700		0.7175		0.4025
0.6125		0.5425		0.3675
}\datatable




\begin{axis}[
   ybar,
    ticks=both,
    axis x line = bottom,
    axis y line = left,
    axis line style={-|},
    nodes near coords = \rotatebox{90}{{\pgfmathprintnumber[fixed zerofill, precision=2]{\pgfplotspointmeta}}},
    nodes near coords align={vertical},
    every node near coord/.append style={font=\small, fill=white, yshift=0.5mm},
    enlarge y limits={lower, value=0.1},
    enlarge y limits={upper, value=0.22},
    ylabel=\bf \large Time (days),
    xlabel=\bf \large Transmission coefficient of the disease,
    xtick=data,
    ymin = 0,
    ymajorgrids,
    xticklabels={ 
        $\beta=0.1$,
        $\beta=0.2$, 
        $\beta=0.4$,
        $\beta=0.6$,
        $\beta=0.8$,
        $\beta=1$},
    legend style={
    anchor=north east, legend columns=1},
    every axis legend/.append style={nodes={right}, inner sep = 0.2cm},
   x tick label style={align=center, 
   },
    enlarge x limits=0.17,
    width=\textwidth,
    height=7cm,
    bar width=0.4cm,
]
\pgfplotsinvokeforeach {0,...,2}{%
    \addplot table [x expr={\coordindex
    },%
    y index=#1] {\datatable};}%
\legend{Susceptibles\hspace*{8pt}, Infected\hspace*{8pt}, Recovered} 
\end{axis}
\end{tikzpicture}
\end{center}
\mbox{}
\caption{Time (days) to reach the limit number of susceptible, infected and recovered individuals for different values of the transmission coefficient of the disease when an antibiotic with a resistance rate of 97.50\% is prescribed}\label{i2}
\end{figure}

\begin{figure}[H]
\begin{center}
\includegraphics[width=\textwidth]{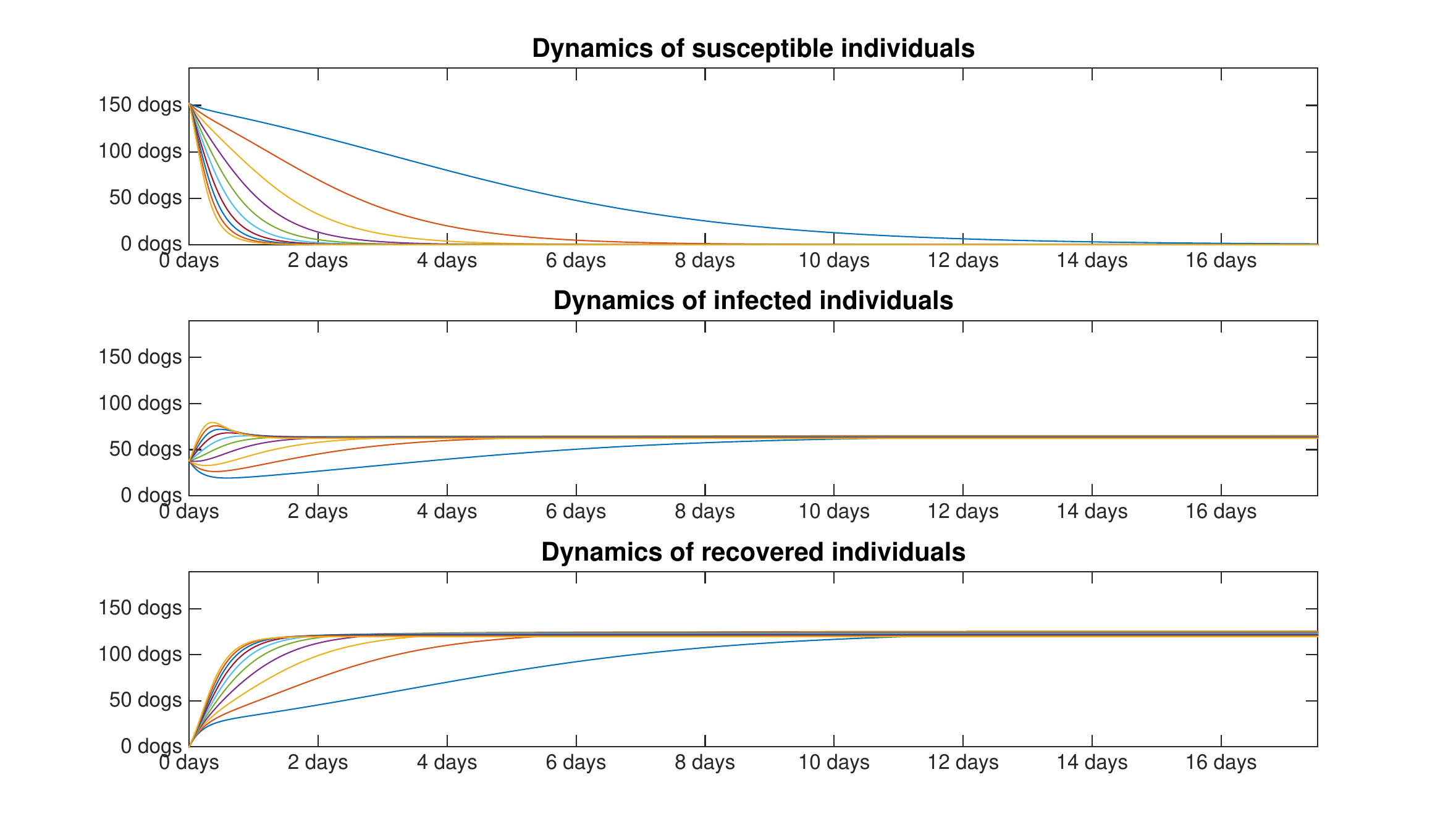}
\end{center}
\caption{Dynamics of susceptible, infected and recovered individuals depending on time (in days) when using an antibiotic with a resistance rate of  34.21\% is prescribed} \label{iii}
\end{figure}

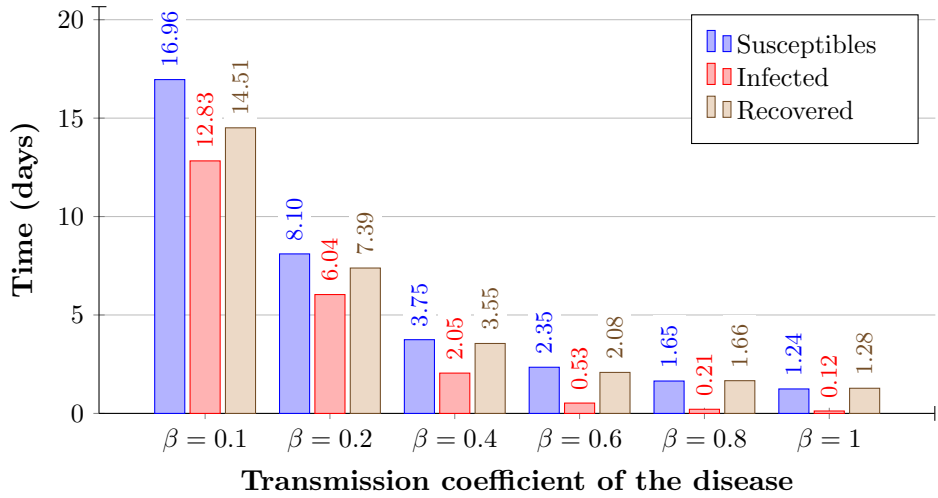
\begin{figure}[H]
\begin{center}

\begin{tikzpicture}
\pgfplotstableread{
16.9575		12.8275		14.5075
8.1025		6.0375		7.3850
3.7450		2.0475		3.5525
2.3450		0.5250		2.0825
1.6450		0.2100		1.6625
1.2425		0.1225		1.2775
}\datatable




\begin{axis}[
   ybar,
    ticks=both,
    axis x line = bottom,
    axis y line = left,
    axis line style={-|},
    nodes near coords = \rotatebox{90}{{\pgfmathprintnumber[fixed zerofill, precision=2]{\pgfplotspointmeta}}},
    nodes near coords align={vertical},
    every node near coord/.append style={font=\small, fill=white, yshift=0.5mm},
    enlarge y limits={lower, value=0.1},
    enlarge y limits={upper, value=0.22},
    ylabel=\bf \large Time (days),
    xlabel=\bf \large Transmission coefficient of the disease,
    xtick=data,
    ymin = 0,
    ymajorgrids,
    xticklabels={ 
        $\beta=0.1$,
        $\beta=0.2$, 
        $\beta=0.4$,
        $\beta=0.6$,
        $\beta=0.8$,
        $\beta=1$},
    legend style={
    anchor=north east, legend columns=1},
    every axis legend/.append style={nodes={right}, inner sep = 0.2cm},
   x tick label style={align=center, 
   },
    enlarge x limits=0.17,
    width=\textwidth,
    height=7cm,
    bar width=0.4cm,
]
\pgfplotsinvokeforeach {0,...,2}{%
    \addplot table [x expr={\coordindex
    },%
    y index=#1] {\datatable};}%
\legend{Susceptibles\hspace*{8pt}, Infected\hspace*{8pt}, Recovered} 
\end{axis}
\end{tikzpicture}
\end{center}
\mbox{}
\caption{Time (days) to reach the limit number of susceptible, infected and recovered individuals for different values of the transmission coefficient of the disease when an antibiotic with a resistance rate of 34.21\% is prescribed}
\label{i3}
\end{figure}

\begin{figure}[H]
\begin{center}
\includegraphics[width=\textwidth]{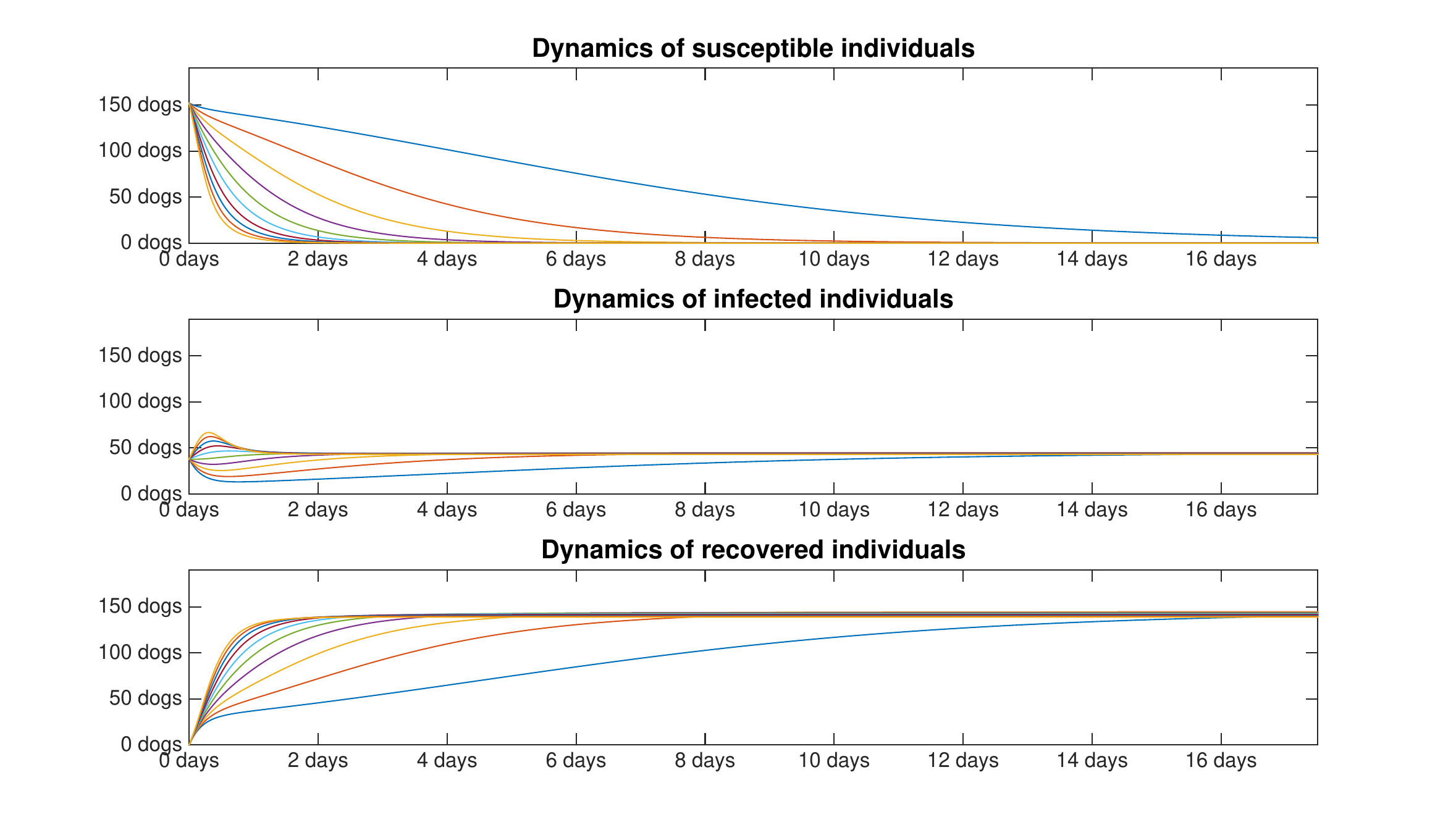}
\end{center}
\caption{Dynamics of susceptible, infected and recovered individuals depending on time (in days) when an antibiotic with a resistance rate of 23.68\% is used}
\label{iv}
\end{figure}

\begin{figure}[H]
\begin{center}

\begin{tikzpicture}
\pgfplotstableread{
17.4475		13.8425		16.5375
11.6025		6.8075		10.2375
5.2675		2.3450		4.6900
3.2200		0.1925		3.0625
2.2400		0.0350		2.0650
1.6625		0.0175		1.5575
}\datatable




\begin{axis}[
   ybar,
    ticks=both,
    axis x line = bottom,
    axis y line = left,
    axis line style={-|},
    nodes near coords = \rotatebox{90}{{\pgfmathprintnumber[fixed zerofill, precision=2]{\pgfplotspointmeta}}},
    nodes near coords align={vertical},
    every node near coord/.append style={font=\small, fill=white, yshift=0.5mm},
    enlarge y limits={lower, value=0.1},
    enlarge y limits={upper, value=0.22},
    ylabel=\bf \large Time (days),
    xlabel=\bf \large Transmission coefficient of the disease,
    xtick=data,
    ymin = 0,
    ymajorgrids,
    xticklabels={ 
        $\beta=0.1$,
        $\beta=0.2$, 
        $\beta=0.4$,
        $\beta=0.6$,
        $\beta=0.8$,
        $\beta=1$},
    legend style={
    anchor=north east, legend columns=1},
    every axis legend/.append style={nodes={right}, inner sep = 0.2cm},
   x tick label style={align=center, 
   },
    enlarge x limits=0.17,
    width=\textwidth,
    height=7cm,
    bar width=0.4cm,
]
\pgfplotsinvokeforeach {0,...,2}{%
    \addplot table [x expr={\coordindex
    },%
    y index=#1] {\datatable};}%
\legend{Susceptibles\hspace*{8pt}, Infected\hspace*{8pt}, Recovered} 
\end{axis}
\end{tikzpicture}
\end{center}
\mbox{}
\caption{Time (days) to reach the limit number of susceptible, infected and recovered individuals for different values of the transmission coefficient of the disease in case of using antibiotics with a resistance rate of  23.68\%}
\label{i4}
\end{figure}
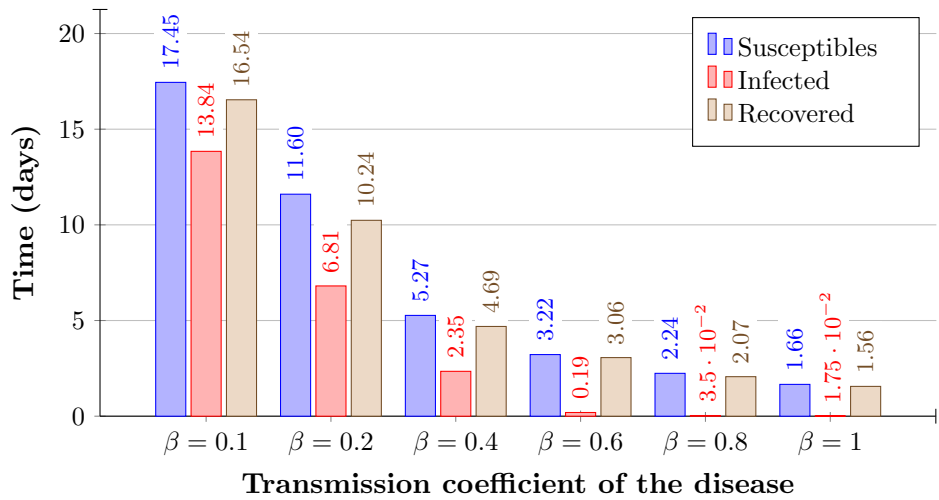

\begin{figure}[H]
\begin{center}
\includegraphics[width=\textwidth]{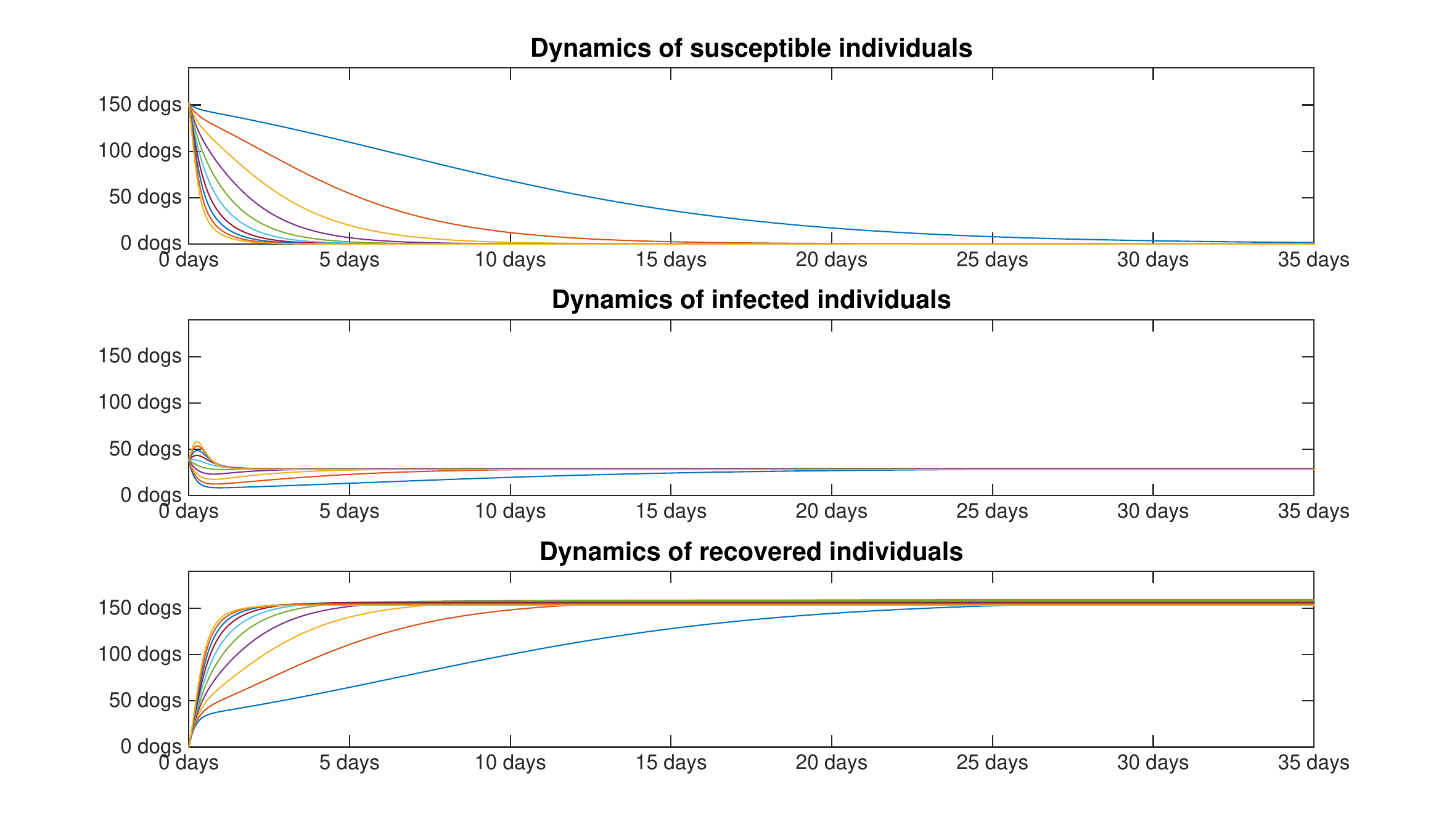}
\end{center}
\caption{Dynamics of susceptible, infected and recovered individuals depending on time (in days) when using an antibiotic with a resistance rate of 15.15\%}
\label{v}
\end{figure}

\begin{figure}[H]
\begin{center}

\begin{tikzpicture}
\pgfplotstableread{
33.2850		23.6075		30.7300
17.6050		10.1500		16.3275
7.8575		3.2025		7.1575
4.7425		2.8350		4.5325
3.2550		2.1175		2.9575
2.3800		1.5750		2.1875

}\datatable




\begin{axis}[
   ybar,
    ticks=both,
    axis x line = bottom,
    axis y line = left,
    axis line style={-|},
    nodes near coords = \rotatebox{90}{{\pgfmathprintnumber[fixed zerofill, precision=2]{\pgfplotspointmeta}}},
    nodes near coords align={vertical},
    every node near coord/.append style={font=\small, fill=white, yshift=0.5mm},
    enlarge y limits={lower, value=0.1},
    enlarge y limits={upper, value=0.22},
    ylabel=\bf \large Time (days),
    xlabel=\bf \large Transmission coefficient of the disease,
    xtick=data,
    ymin = 0,
    ymajorgrids,
    xticklabels={ 
        $\beta=0.1$,
        $\beta=0.2$,
        $\beta=0.2$, 
        $\beta=0.4$,
        $\beta=0.6$,
        $\beta=0.8$,
        $\beta=1$},
    legend style={
    anchor=north east, legend columns=1},
    every axis legend/.append style={nodes={right}, inner sep = 0.2cm},
   x tick label style={align=center, 
   },
    enlarge x limits=0.17,
    width=\textwidth,
    height=7cm,
    bar width=0.4cm,
]
\pgfplotsinvokeforeach {0,...,2}{%
    \addplot table [x expr={\coordindex
    },%
    y index=#1] {\datatable};}%
\legend{Susceptibles\hspace*{8pt}, Infected\hspace*{8pt}, Recovered} 
\end{axis}
\end{tikzpicture}
\end{center}
\mbox{}
\caption{Time (days) to reach the limit number of susceptible, infected and recovered individuals for different values of the transmission coefficient of the disease in case when an antibiotic with resistance rate of 15.15\% is used}
\label{i5}
\end{figure}
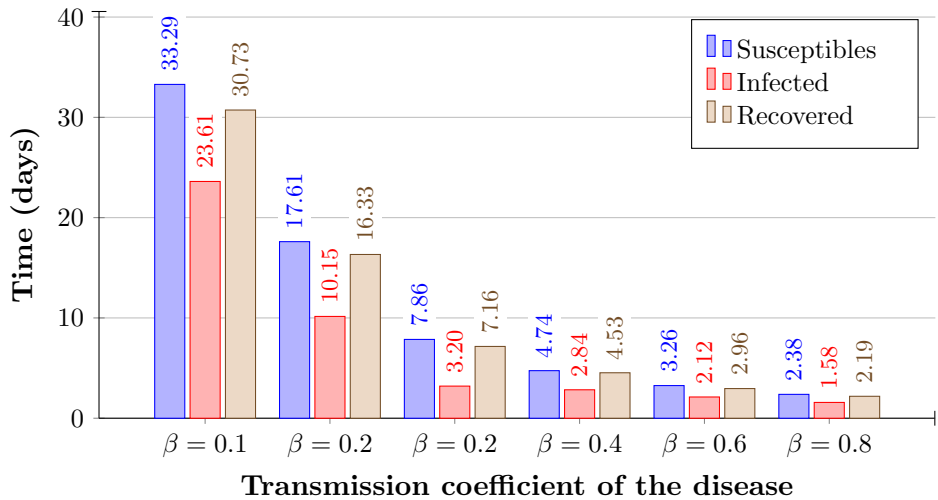

\begin{figure}[H]
\begin{center}
\includegraphics[width=\textwidth]{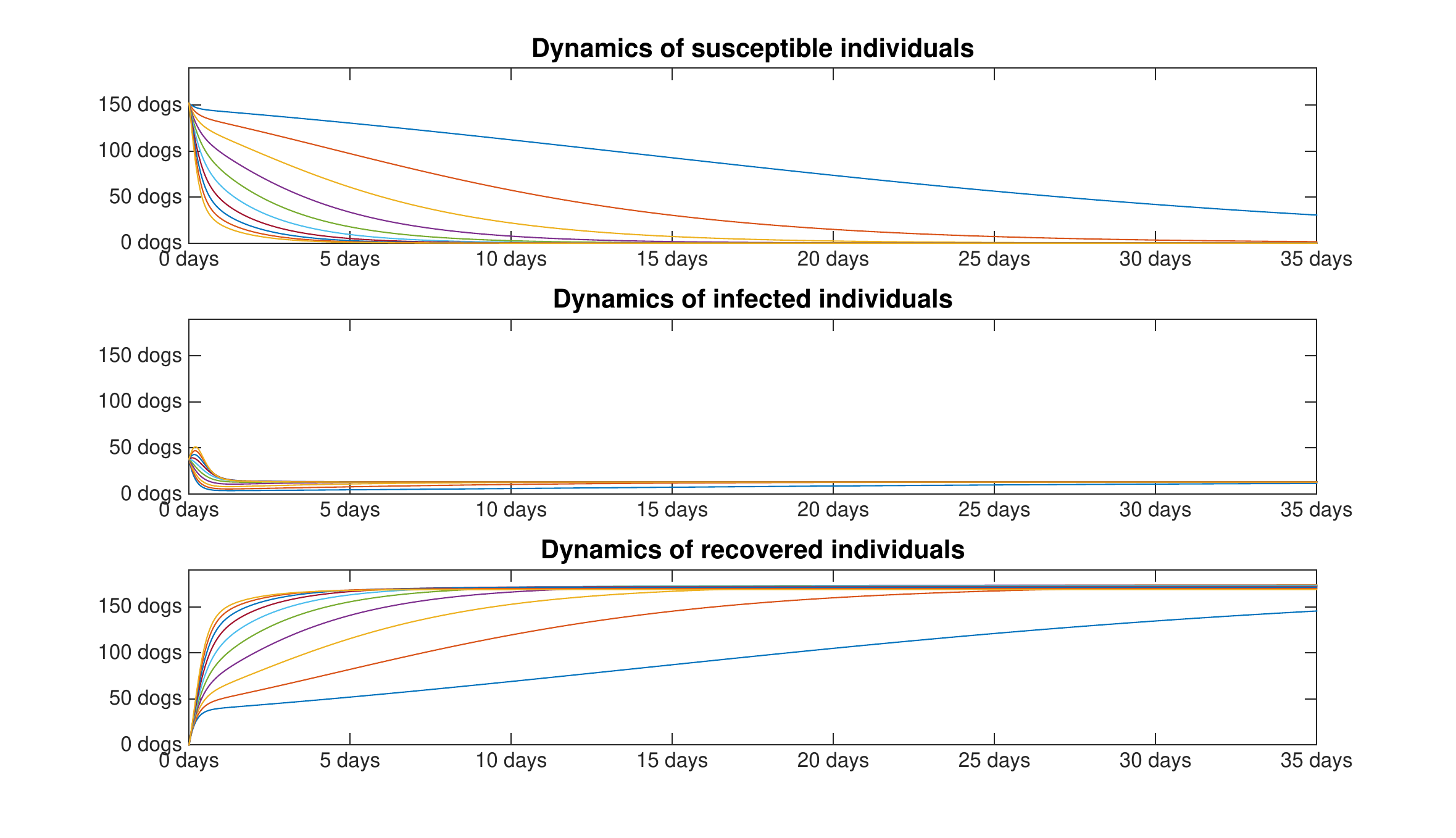}
\end{center}
\caption{Dynamics of susceptible, infected and recovered individuals depending on time (in days) when an antibiotic with a resistance rate of  7.14\% is assayed}
\label{vi}
\end{figure}

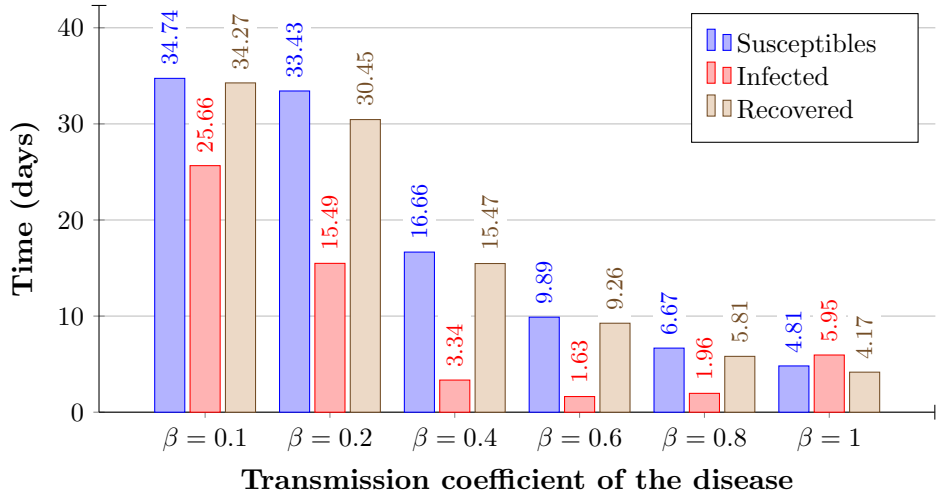
\begin{figure}[H]
\begin{center}

\begin{tikzpicture}
\pgfplotstableread{
34.7375		25.6550		34.2650
33.4250		15.4875		30.4500
16.6600		3.3425		15.4700
9.8875		1.6275		9.2575
6.6675		1.9600		5.8100
4.8125		5.9500		4.1650
}\datatable




\begin{axis}[
   ybar,
    ticks=both,
    axis x line = bottom,
    axis y line = left,
    axis line style={-|},
    nodes near coords = \rotatebox{90}{{\pgfmathprintnumber[fixed zerofill, precision=2]{\pgfplotspointmeta}}},
    nodes near coords align={vertical},
    every node near coord/.append style={font=\small, fill=white, yshift=0.5mm},
    enlarge y limits={lower, value=0.1},
    enlarge y limits={upper, value=0.22},
    ylabel=\bf \large Time (days),
    xlabel=\bf \large Transmission coefficient of the disease,
    xtick=data,
    ymin = 0,
    ymajorgrids,
    xticklabels={ 
        $\beta=0.1$,
        $\beta=0.2$, 
        $\beta=0.4$,
        $\beta=0.6$,
        $\beta=0.8$,
        $\beta=1$},
    legend style={
    anchor=north east, legend columns=1},
    every axis legend/.append style={nodes={right}, inner sep = 0.2cm},
   x tick label style={align=center, 
   },
    enlarge x limits=0.17,
    width=\textwidth,
    height=7cm,
    bar width=0.4cm,
]
\pgfplotsinvokeforeach {0,...,2}{%
    \addplot table [x expr={\coordindex
    },%
    y index=#1] {\datatable};}%
\legend{Susceptibles\hspace*{8pt}, Infected\hspace*{8pt}, Recovered} 
\end{axis}
\end{tikzpicture}
\end{center}
\mbox{}
\caption{Time (days) to reach the limit number of susceptible, infected and recovered individuals for different values of the transmission coefficient of the disease in case of using antibiotics with a resistance rate of 7.14\%}
\label{i6}
\end{figure}

\begin{figure}[H]
\begin{center}
\includegraphics[width=\textwidth]{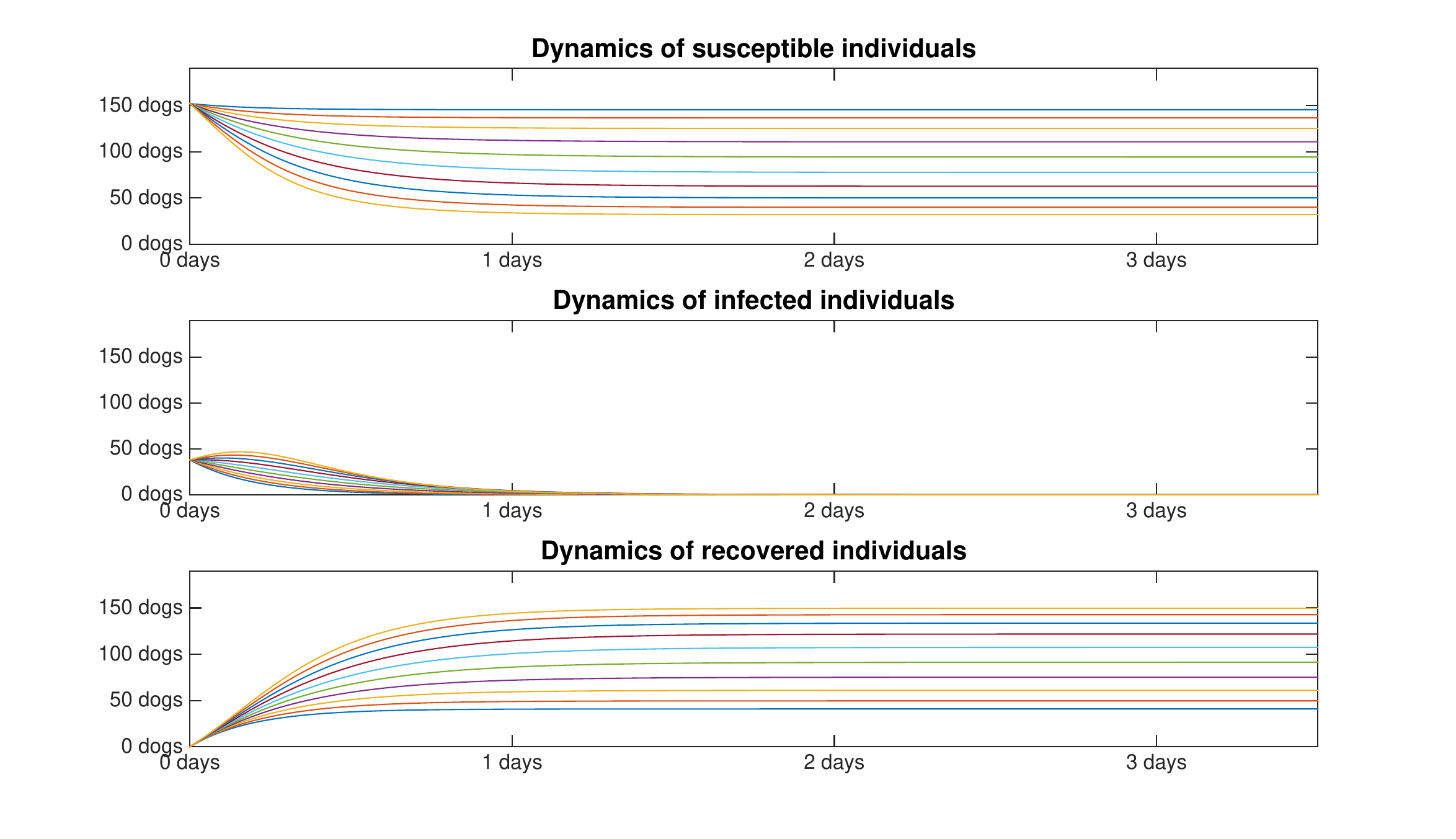}
\end{center}
\caption{Dynamics of susceptible, infected and recovered individuals depending on time (in days) when an antibiotic with a resistance rate of 0\% is used}
\label{vii}
\end{figure}

\begin{figure}[H]
\begin{center}

\begin{tikzpicture}
\pgfplotstableread{
0.4375		0.7000		0.5950
0.8225		0.8400		0.8050
1.4875		1.1900		1.3125
1.6800		1.4175		1.4875
2.6425		1.4175		1.4175
1.6975		1.3300		1.2775
}\datatable




\begin{axis}[
   ybar,
    ticks=both,
    axis x line = bottom,
    axis y line = left,
    axis line style={-|},
    nodes near coords = \rotatebox{90}{{\pgfmathprintnumber[fixed zerofill, precision=2]{\pgfplotspointmeta}}},
    nodes near coords align={vertical},
    every node near coord/.append style={font=\small, fill=white, yshift=0.5mm},
    enlarge y limits={lower, value=0.1},
    enlarge y limits={upper, value=0.22},
    ylabel=\bf \large Time (days),
    xlabel=\bf \large Transmission coefficient of the disease,
    xtick=data,
    ymin = 0,
    ymajorgrids,
    xticklabels={ 
        $\beta=0.1$,
        $\beta=0.2$, 
        $\beta=0.4$,
        $\beta=0.6$,
        $\beta=0.8$,
        $\beta=1$},
    legend style={
    anchor=north east, legend columns=1},
    every axis legend/.append style={nodes={right}, inner sep = 0.2cm},
   x tick label style={align=center, 
   },
    enlarge x limits=0.17,
    width=\textwidth,
    height=7cm,
    bar width=0.4cm,
]
\pgfplotsinvokeforeach {0,...,2}{%
    \addplot table [x expr={\coordindex
    },%
    y index=#1] {\datatable};}%
\legend{Susceptibles\hspace*{8pt}, Infected\hspace*{8pt}, Recovered} 
\end{axis}
\end{tikzpicture}
\end{center}
\mbox{}
\caption{Time (days) to reach the limit number of susceptible, infected and recovered individuals for different values of the transmission coefficient of the disease in case of using antibiotics with a resistance rate of 0\%}
\label{i7}
\end{figure}
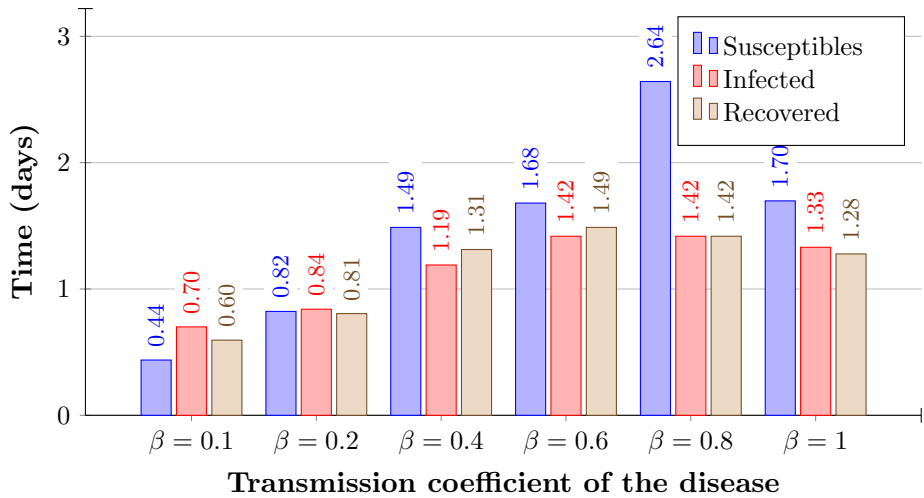

For the implementation of the model, three tools are considered essential: 

\begin{itemize}
\item	Figures (Figures \ref{i}, \ref{2}, \ref{iii}, \ref{iv}, \ref{v}, \ref{vi}, \ref{vii}) containing numerical simulations of the SIRI model \eqref{1}-\eqref{3} with a code developed in {\it Matlab}, which will allow to determine the dynamics of the susceptible, infected and recovered individuals for an estimated period of time. Moreover, it allows observing how the numbers of the different groups of our population change and its asymptotic behavior.
\item	Tables (Tables \ref{d1}, \ref{d2}, \ref{d3}, \ref{d4}, \ref{d5}, \ref{d6}, \ref{d7}) where the limit number of each group of the population is collected for different values of the transmission coefficient of the disease, as well as the time to reach the corresponding value. These tables are very useful to convert the dynamics information in numerical data. This second tool was implemented due to a code programmed in {\it Matlab}.
\item Bar diagrams (Figures \ref{i1}, \ref{i2}, \ref{i3}, \ref{i4}, \ref{i5}, \ref{i6}, \ref{i7}) where the final number of each group of the population is presented for different values of the transmission coefficient of the disease.
\end{itemize}

Once the mathematical study is accomplished, it could provide practitioners an appropriate and very convenient panel of the most effective antibiotics, which will provide essential information in case the antibiogram is not available.\n

In order to develop the mathematical study, a population of 192 dogs were included, where 40 of them are infected by {\it Pseudomonas aeruginosa}, 152 are susceptible and none of them are recovered at the initial time, considering this initial value as zero. \n

Concerning the parameters in the SIRI model \eqref{1}-\eqref{3} the birth and death rates of the population were $\Lambda=0$ and $\mu=0$, assuming that no dogs were born or died during our study. Since the average duration of this infection is considered around 14 weeks, $\alpha=0$ since dog mortality null in this pathology. The recovery rate $\kappa$ and the reinfection rate $\gamma$ depend on the groups of antibiotics in study (see Section \ref{work}). We remark that the transmission coefficient of the disease is unknown but it is a rate. Therefore, every case will be studied using different values of this parameter between 0.1 and 1.

\subsection{Resistance 100\%}

This subsection corresponds to antibiotics with a resistance rate of 100\%.\n

\begin{figure}[H]
\begin{center}
\includegraphics[width=\textwidth]{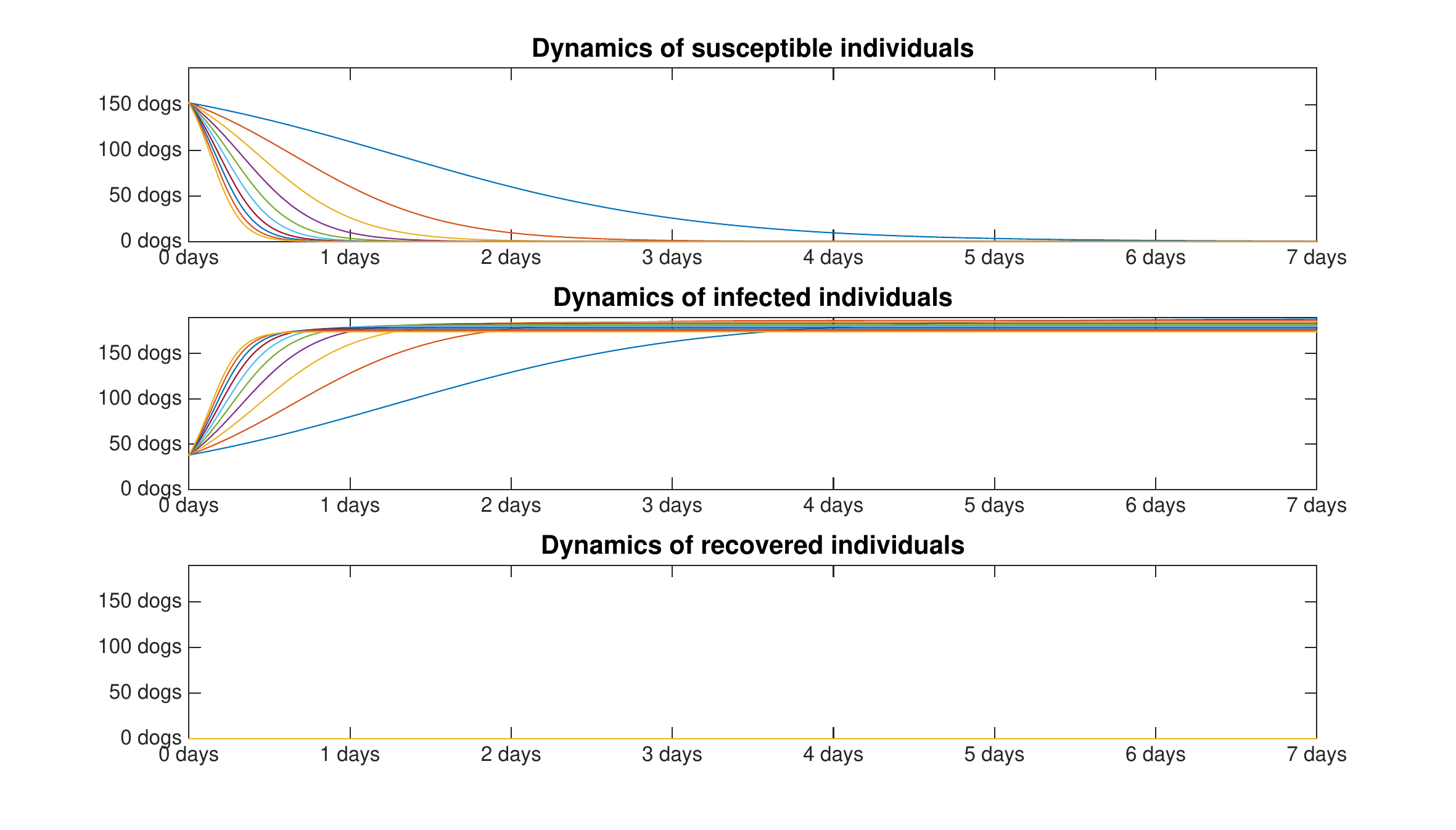}
\end{center}
\caption{Dynamics of susceptible, infected and recovered individuals depending on time (in days) when using an antibiotic with a resistance rate of 100\%}\label{i}
\end{figure}

In Figure \ref{i} we display three different panels showing the dynamics of susceptible, infected and recovered individuals, respectively. In every panel the number of members in each group are plotted depending on time. As the initial number of dogs in the experiment is 192 and only 40 are infected, the initial value of susceptible is 152 and the initial value of infected as 40. Noticeable, there are no recovered dogs at the initial time.

In Table \ref{d1} the number of susceptible, infected and recovered individuals at the final time is depicted for different values of the transmission coefficient of the disease $\beta$. For instance, when $\beta=0.1$ the limit of susceptible individuals is 0.9892 and this limit is reached after 6.1075 days. The limit of infected individuals is 186.0153 and it is reached after 5.3200 days and, finally, the number of susceptible individuals remains constant zero during all the time.\n

In Table \ref{d1} we can notice that most of dogs will be infected in approximately 5 days and no dogs will be recovered corresponding to a 100\% resistance rate.

\begin{table}[H]
\begin{center}
\resizebox{\textwidth}{!}{ 
\begin{tabular}{ccccccc}
\hline
\multicolumn{ 1}{c}{} & \multicolumn{ 2}{c}{Susceptibles} & \multicolumn{ 2}{c}{Infected} & \multicolumn{ 2}{c}{Recovered} \\ \cline{ 2- 7}
\multicolumn{ 1}{c}{} & Time (days) & Individuals & Time (days) & Individuals & Time (days) & Individuals \\ \hline
\rowcolor[gray]{0.9} $\beta=0.1$ & 6.1075 & 0.9892 & 5.3200 & 186.0153 & 0.0000 & 0.0000 \\ 
$\beta=0.2$ & 3.0450 & 0.9997 & 2.8175 & 185.0218 & 0.0000 & 0.0000 \\ 
\rowcolor[gray]{0.9} $\beta=0.3$ & 2.0300 & 0.9911 & 1.8200 & 183.0880 & 0.0000 & 0.0000 \\ 
$\beta=0.4$ & 1.5225 & 0.9821 & 1.4525 & 182.0727 & 0.0000 & 0.0000 \\ 
\rowcolor[gray]{0.9} $\beta=0.5$ & 1.2250 & 0.9368 & 1.1025 & 180.0496 & 0.0000 & 0.0000 \\ 
$\beta=0.6$ & 1.0150 & 0.9635 & 0.9800 & 179.0448 & 0.0000 & 0.0000 \\ 
\rowcolor[gray]{0.9} $\beta=0.7$ & 0.8750 & 0.9183 & 0.7875 & 177.0125 & 0.0000 & 0.0000 \\ 
$\beta=0.8$ & 0.7700 & 0.8748 & 0.7350 & 176.0281 & 0.0000 & 0.0000 \\ 
\rowcolor[gray]{0.9} $\beta=0.9$ & 0.6825 & 0.8819 & 0.6125 & 174.0724 & 0.0000 & 0.0000 \\ 
$\beta=1$ & 0.6125 & 0.8888 & 0.5950 & 173.1766 & 0.0000 & 0.0000 \\ \hline
\end{tabular}
}
\end{center}
\caption{Data for different values of the transmission coefficient of the disease $\beta$ in case of using antibiotics with a resistance rate of 100\%}\label{d1}
\end{table}

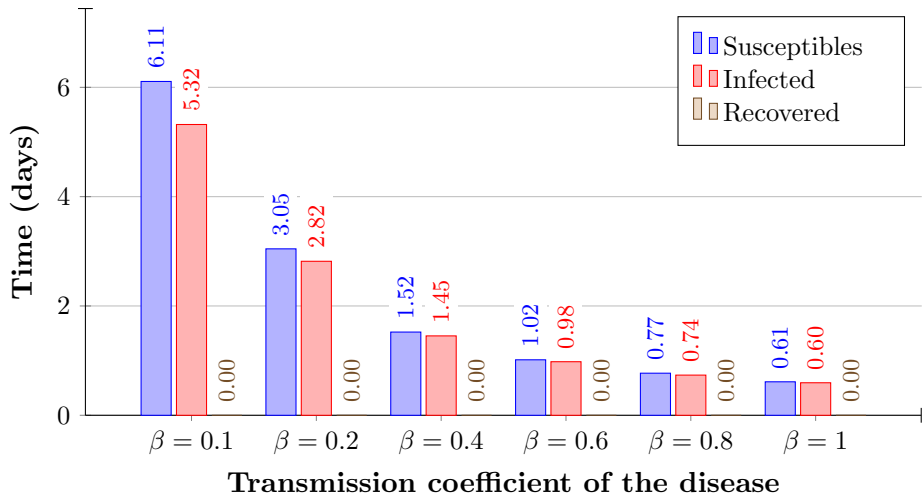
\begin{figure}[H]
\begin{center}
\begin{tikzpicture}
\pgfplotstableread{
6.1075		5.3200		0.0000
3.0450		2.8175		0.0000
1.5225		1.4525		0.0000
1.0150		0.9800		0.0000
0.7700		0.7350		0.0000
0.6125		0.5950		0.0000

}\datatable




\begin{axis}[
   ybar,
    ticks=both,
    axis x line = bottom,
    axis y line = left,
    axis line style={-|},
    nodes near coords = \rotatebox{90}{{\pgfmathprintnumber[fixed zerofill, precision=2]{\pgfplotspointmeta}}},
    nodes near coords align={vertical},
    every node near coord/.append style={font=\small, fill=white, yshift=0.5mm},
    enlarge y limits={lower, value=0.1},
    enlarge y limits={upper, value=0.22},
    ylabel=\bf \large Time (days),
    xlabel=\bf \large Transmission coefficient of the disease,
    xtick=data,
    ymin = 0,
    ymajorgrids,
    xticklabels={
        $\beta=0.1$, 
        $\beta=0.2$, 
        $\beta=0.4$,
        $\beta=0.6$,
        $\beta=0.8$,
        $\beta=1$},
    legend style={
    anchor=north east, legend columns=1},
    every axis legend/.append style={nodes={right}, inner sep = 0.2cm},
   x tick label style={align=center, 
   },
    enlarge x limits=0.17,
    width=\textwidth,
    height=7cm,
    bar width=0.4cm,
]
\pgfplotsinvokeforeach {0,...,2}{%
    \addplot table [x expr={\coordindex
    },%
    y index=#1] {\datatable};}%
\legend{Susceptibles\hspace*{8pt}, Infected\hspace*{8pt}, Recovered} 
\end{axis}
\end{tikzpicture}
\end{center}
\mbox{}
\caption{Time (days) to reach the limit number of susceptible, infected and recovered individuals for different values of the transmission coefficient of the disease in case of using antibiotics with a
resistance rate of 100\%}\label{i1}
\end{figure}

Based on Table \ref{d1} we present in Figure \ref{i1} a bar diagram to to illustrate the results obtained.

In order to avoid redundancies, in the following cases we will present the different figures following the structure presented before.

\subsection{Resistance close to 100\%}

This subsection corresponds to antibiotics with a resistance rate close to 100\%. Due to the similarities of the rates among the antibiotics of this group, Amoxicillin and Clavulanic acid has been considered to be representative, with a resistance rate of 97.50\%.

\begin{figure}[H]
\begin{center}
\includegraphics[width=\textwidth]{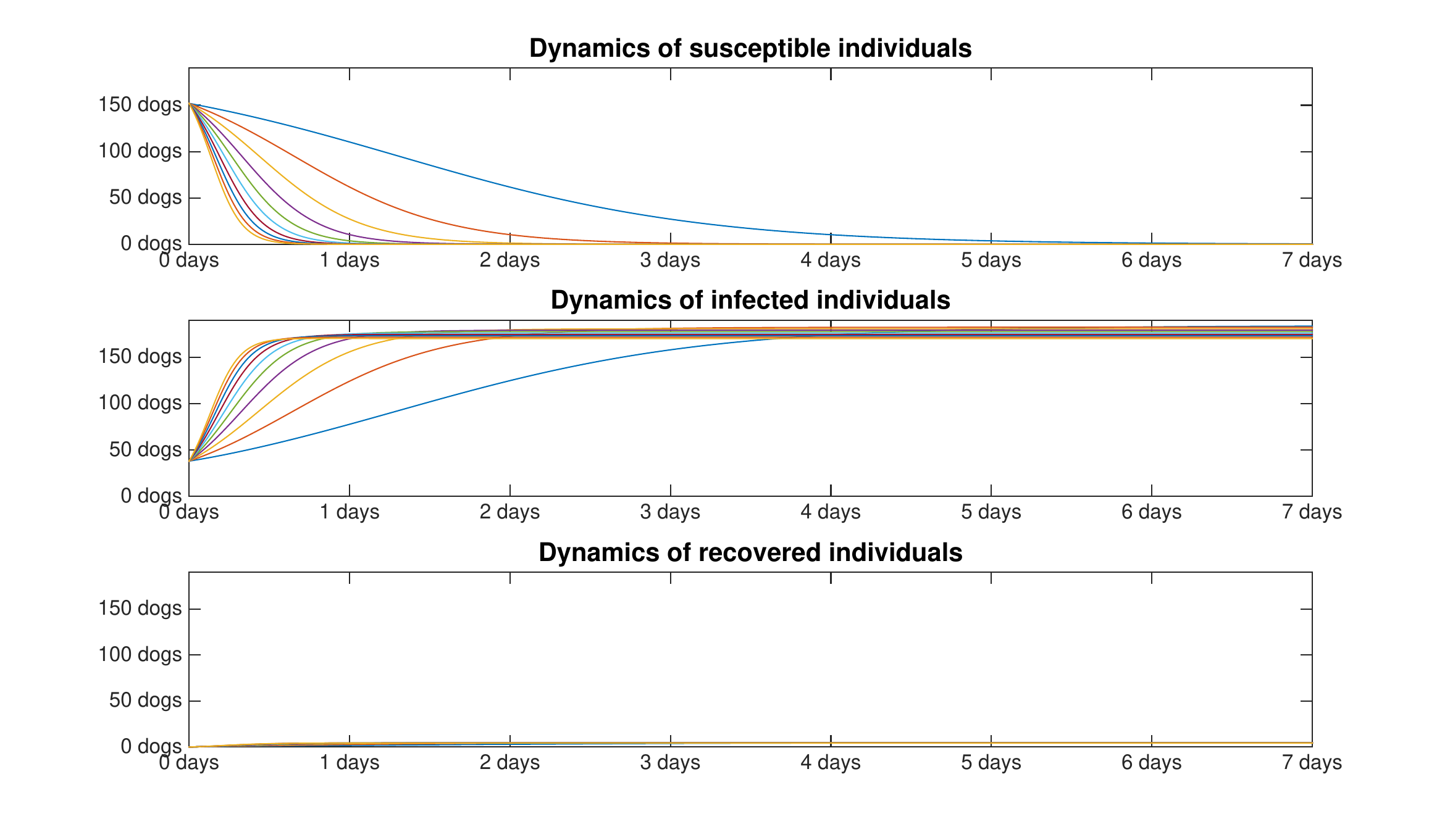}
\end{center}
\caption{Dynamics of susceptible, infected and recovered individuals depending on time (in days) when using an antibiotic with resistance 97.50\%}\label{2}
\end{figure}

\begin{table}[H]
\begin{center}
\resizebox{\textwidth}{!}{ 
\begin{tabular}{ccccccc}
\hline
\multicolumn{ 1}{c}{} & \multicolumn{ 2}{c}{Susceptibles} & \multicolumn{ 2}{c}{Infected} & \multicolumn{ 2}{c}{Recovered} \\ \cline{ 2- 7}
\multicolumn{ 1}{c}{} & Time (days) & Individuals & Time (days) & Individuals & Time (days) & Individuals \\ \hline
\rowcolor[gray]{0.9} $\beta=0.1$ & 6.2475 & 0.9857 & 5.5475 & 182.0012 & 1.9775 & 3.0085 \\ 
$\beta=0.2$ & 3.1150 & 0.9875 & 2.9400 & 181.0199 & 1.0675 & 3.0005 \\ 
\rowcolor[gray]{0.9}$\beta=0.3$ & 2.0825 & 0.9541 & 1.8725 & 179.0799 & 0.7700 & 3.0049 \\ 
$\beta=0.4$ & 1.5575 & 0.9571 & 1.4875 & 178.0629 & 0.6300 & 3.0395 \\ 
\rowcolor[gray]{0.9}$\beta=0.5$ & 1.2425 & 0.9603 & 1.1200 & 176.1075 & 0.5425 & 3.0513 \\ 
$\beta=0.6$ & 1.0325 & 0.9637 & 0.9800 & 175.0559 & 0.4900 & 3.0900 \\ 
\rowcolor[gray]{0.9}$\beta=0.7$ & 0.8925 & 0.8982 & 0.7875 & 173.1549 & 0.4375 & 3.0372 \\ 
$\beta=0.8$ & 0.7700 & 0.9709 & 0.7175 & 172.1004 & 0.4025 & 3.0154 \\ 
\rowcolor[gray]{0.9}$\beta=0.9$ & 0.6825 & 0.9746 & 0.5950 & 170.2078 & 0.3850 & 3.0559 \\ 
$\beta=1$ & 0.6125 & 0.9783 & 0.5425 & 169.0158 & 0.3675 & 3.0651 \\ \hline
\end{tabular}
}
\end{center}
\caption{Data for different values of the transmission coefficient of the disease $\beta$ in case of using antibiotics with a resistance rate of 97.50\%}\label{d2}
\end{table}

\subsection{Resistance 30\%-50\%}

This subsection corresponds to the group of antibiotics with a resistance rate between 30\% and 50\%. Among all the antibiotics included, Enrofloxacin, with a resistance rate of 34,21\%, is used as an example.

\begin{table}[H]
\begin{center}
\resizebox{\textwidth}{!}{ 
\begin{tabular}{ccccccc}
\hline
\multicolumn{ 1}{c}{} & \multicolumn{ 2}{c}{Susceptibles} & \multicolumn{ 2}{c}{Infected} & \multicolumn{ 2}{c}{Recovered} \\ \cline{ 2- 7}
\multicolumn{ 1}{c}{} & Time (days) & Individuals & Time (days) & Individuals & Time (days) & Individuals \\ \hline
\rowcolor[gray]{0.9} $\beta=0.1$ & 16.9575 & 0.9994 & 12.8275 & 64.0028 & 14.5075 & 124.0053 \\ 
$\beta=0.2$ & 8.1025 & 0.9991 & 6.0375 & 64.0011 & 7.3850 & 124.0110 \\ 
\rowcolor[gray]{0.9} $\beta=0.3$ & 5.1975 & 0.9824 & 3.1325 & 63.0067 & 4.4975 & 123.0316 \\ 
$\beta=0.4$ & 3.7450 & 0.9942 & 2.0475 & 63.0236 & 3.5525 & 123.0152 \\ 
\rowcolor[gray]{0.9} $\beta=0.5$ & 2.9050 & 0.9703 & 1.2250 & 63.0482 & 2.6425 & 122.0149 \\ 
$\beta=0.6$ & 2.3450 & 0.9633 & 0.5250 & 62.2112 & 2.0825 & 121.0022 \\ 
\rowcolor[gray]{0.9} $\beta=0.7$ & 1.9425 & 0.9752 & 0.3150 & 62.8271 & 1.9775 & 121.0016 \\ 
$\beta=0.8$ & 1.6450 & 0.9842 & 0.2100 & 62.7526 & 1.6625 & 120.0108 \\ 
\rowcolor[gray]{0.9} $\beta=0.9$ & 1.4175 & 0.9882 & 1.5225 & 62.9793 & 1.4525 & 119.0726 \\ 
$\beta=1$ & 1.2425 & 0.9731 & 0.1225 & 62.8659 & 1.2775 & 118.0144 \\ \hline
\end{tabular}
}
\end{center}
\caption{Data for different values of the transmission coefficient of the disease $\beta$ when an antibiotic with a resistance rate of 34.21\% is used}
\label{d3}
\end{table}

\subsection{Resistance 20\%-30\%}

This subsection is focused in the group of antibiotics with a resistance rate between 20 and 30 Gentamicin, with a resistance rate of 23.68\%, has been chosen as an example.

\begin{table}[H]
\begin{center}
\resizebox{\textwidth}{!}{ 
\begin{tabular}{ccccccc}
\hline
\multicolumn{ 1}{c}{} & \multicolumn{ 2}{c}{Susceptibles} & \multicolumn{ 2}{c}{Infected} & \multicolumn{ 2}{c}{Recovered} \\ \cline{ 2- 7}
\multicolumn{ 1}{c}{} & Time (days) & Individuals & Time (days) & Individuals & Time (days) & Individuals \\ \hline
\rowcolor[gray]{0.9} $\beta=0.1$ & 17.4475 & 5.9890 & 13.8425 & 42.0097 & 16.5375 & 139.0219 \\ 
$\beta=0.2$ & 11.6025 & 0.9932 & 6.8075 & 43.0015 & 10.2375 & 143.0048 \\ 
\rowcolor[gray]{0.9}$\beta=0.3$ & 7.3500 & 0.9962 & 3.9025 & 43.0159 & 6.9300 & 143.0074 \\ 
$\beta=0.4$ & 5.2675 & 0.9891 & 2.3450 & 43.0155 & 4.6900 & 142.0205 \\ 
\rowcolor[gray]{0.9}$\beta=0.5$ & 4.0425 & 0.9787 & 1.1025 & 43.0038 & 3.9900 & 142.0091 \\ 
$\beta=0.6$ & 3.2200 & 0.9987 & 0.1925 & 43.2377 & 3.0625 & 141.0019 \\ 
\rowcolor[gray]{0.9}$\beta=0.7$ & 2.6600 & 0.9864 & 0.0525 & 42.0490 & 2.4675 & 140.0001 \\ 
$\beta=0.8$ & 2.2400 & 0.9866 & 0.0350 & 42.6969 & 2.0650 & 139.0233 \\ 
\rowcolor[gray]{0.9}$\beta=0.9$ & 1.9250 & 0.9699 & 0.0175 & 42.2403 & 1.7675 & 138.0169 \\ 
$\beta=1$ & 1.6625 & 0.9865 & 0.0175 & 43.3302 & 1.5575 & 137.0543 \\ \hline
\end{tabular}
}
\end{center}
\caption{Data for different values of the transmission coefficient of the disease $\beta$ when an antibiotic with a resistance rate of 23.68\% is evaluated}
\label{d4}
\end{table}

\subsection{Resistance 10\%-20\%}

In this subsection the antibiotics with a resistance rate between 10\% and 20\% are studied. In this case, Tobramycin, with a resistance rate of 15,15\% has been considered as an example.

\begin{table}[H]
\begin{center}
\resizebox{\textwidth}{!}{ 
\begin{tabular}{ccccccc}
\hline
\multicolumn{ 1}{c}{} & \multicolumn{ 2}{c}{Susceptibles} & \multicolumn{ 2}{c}{Infected} & \multicolumn{ 2}{c}{Recovered} \\ \cline{ 2- 7}
\multicolumn{ 1}{c}{} & Time (days) & Individuals & Time (days) & Individuals & Time (days) & Individuals \\ \hline
\rowcolor[gray]{0.9} $\beta=0.1$ & 33.2850 & 1.9970 & 23.6075 & 28.0021 & 30.7300 & 157.0074 \\ 
$\beta=0.2$ & 17.6050 & 0.9952 & 10.1500 & 28.0028 & 16.3275 & 158.0060 \\ 
\rowcolor[gray]{0.9}$\beta=0.3$ & 11.0775 & 0.9917 & 5.6175 & 28.0040 & 9.5200 & 157.0011 \\ 
$\beta=0.4$ & 7.8575 & 0.9982 & 3.2025 & 28.0095 & 7.1575 & 157.0100 \\ 
\rowcolor[gray]{0.9}$\beta=0.5$ & 5.9850 & 0.9881 & 0.6300 & 28.9433 & 5.1625 & 156.0244 \\ 
$\beta=0.6$ & 4.7425 & 0.9984 & 2.8350 & 28.9999 & 4.5325 & 156.0066 \\ 
\rowcolor[gray]{0.9}$\beta=0.7$ & 3.8850 & 0.9942 & 2.5200 & 28.9953 & 3.6050 & 155.0242 \\ 
$\beta=0.8$ & 3.2550 & 0.9880 & 2.1175 & 28.9973 & 2.9575 & 154.0025 \\ 
\rowcolor[gray]{0.9}$\beta=0.9$ & 2.7650 & 0.9933 & 1.8025 & 28.9944 & 2.5200 & 153.0369 \\ 
$\beta=1$ & 2.3800 & 0.9956 & 1.5750 & 28.9719 & 2.1875 & 152.0432 \\ \hline
\end{tabular}
}
\end{center}
\caption{Data for different values of the transmission coefficient of the disease $\beta$ in case of using antibiotics with a resistance rate of 15.15\%}
\label{d5}
\end{table}

\subsection{Resistance 1\%-10\%}

This subsection studies the group of antibiotics with a resistance rate between 1\% and 10\%. 
Imipenem, with a resistance rate of 7,14\% is considered representative.

\begin{table}[H]
\begin{center}
\resizebox{\textwidth}{!}{ 
\begin{tabular}{ccccccc}
\hline
\multicolumn{ 1}{c}{} & \multicolumn{ 2}{c}{Susceptibles} & \multicolumn{ 2}{c}{Infected} & \multicolumn{ 2}{c}{Recovered} \\ \cline{ 2- 7}
\multicolumn{ 1}{c}{} & Time (days) & Individuals & Time (days) & Individuals & Time (days) & Individuals \\ \hline
\rowcolor[gray]{0.9} $\beta=0.1$ & 34.7375 & 30.9964 & 25.6550 & 10.0024 & 34.2650 & 144.0183 \\ 
$\beta=0.2$ & 33.4250 & 1.9991 & 15.4875 & 12.0004 & 30.4500 & 171.0015 \\ 
\rowcolor[gray]{0.9}$\beta=0.3$ & 23.6775 & 0.9966 & 7.5425 & 12.0045 & 21.0700 & 172.0027 \\ 
$\beta=0.4$ & 16.6600 & 0.9966 & 3.3425 & 12.0015 & 15.4700 & 172.0035 \\ 
\rowcolor[gray]{0.9}$\beta=0.5$ & 12.5475 & 0.9978 & 0.9975 & 13.9876 & 10.7975 & 171.0122 \\ 
$\beta=0.6$ & 9.8875 & 0.9940 & 1.6275 & 13.9938 & 9.2575 & 171.0008 \\ 
\rowcolor[gray]{0.9}$\beta=0.7$ & 8.0325 & 0.9916 & 1.9950 & 13.9923 & 7.1750 & 170.0007 \\ 
$\beta=0.8$ & 6.6675 & 0.9927 & 1.9600 & 13.9912 & 5.8100 & 169.0100 \\ 
\rowcolor[gray]{0.9}$\beta=0.9$ & 5.6350 & 0.9884 & 1.8025 & 13.9858 & 4.8475 & 168.0026 \\ 
$\beta=1$ & 4.8125 & 0.9920 & 5.9500 & 12.9997 & 4.1650 & 167.0077 \\ \hline
\end{tabular}
}
\end{center}
\caption{Data for different values of the transmission coefficient of the disease $\beta$ in case of using antibiotics with a resistance rate of 7.14\%}
\label{d6}
\end{table}

\subsection{Resistance 0\%}

This subsection corresponds to the antibiotics with total effectiveness.

\begin{table}[H]
\begin{center}
\resizebox{\textwidth}{!}{ 
\begin{tabular}{ccccccc}
\hline
\multicolumn{ 1}{c}{} & \multicolumn{ 2}{c}{Susceptibles} & \multicolumn{ 2}{c}{Infected} & \multicolumn{ 2}{c}{Recovered} \\ \cline{ 2- 7}
\multicolumn{ 1}{c}{} & Time (days) & Individuals & Time (days) & Individuals & Time (days) & Individuals \\ \hline
\rowcolor[gray]{0.9} $\beta=0.1$ & 0.4375 & 145.9988 & 0.7000 & 0.9824 & 0.5950 & 39.0306 \\ 
$\beta=0.2$ & 0.8225 & 136.9863 & 0.8400 & 0.9809 & 0.8050 & 48.0938 \\ 
\rowcolor[gray]{0.9}$\beta=0.3$ & 0.9100 & 125.9459 & 1.0150 & 0.9564 & 0.9450 & 59.0773 \\ 
$\beta=0.4$ & 1.4875 & 110.9878 & 1.1900 & 0.9695 & 1.3125 & 74.0443 \\ 
\rowcolor[gray]{0.9}$\beta=0.5$ & 1.3825 & 94.9995 & 1.3300 & 0.9970 & 1.4525 & 90.0654 \\ 
$\beta=0.6$ & 1.6800 & 77.9836 & 1.4175 & 0.9886 & 1.4875 & 106.0461 \\ 
\rowcolor[gray]{0.9}$\beta=0.7$ & 1.6275 & 62.9900 & 1.4350 & 0.9982 & 1.4175 & 120.0633 \\ 
$\beta=0.8$ & 2.6425 & 49.9998 & 1.4175 & 0.9869 & 1.4175 & 132.0735 \\ 
\rowcolor[gray]{0.9}$\beta=0.9$ & 1.7850 & 39.9951 & 1.3825 & 0.9673 & 1.3300 & 141.0661 \\ 
$\beta=1$ & 1.6975 & 31.9943 & 1.3300 & 0.9899 & 1.2775 & 148.0393 \\ \hline
\end{tabular}
}
\end{center}
\caption{Data for different values of the transmission coefficient of the disease $\beta$ in case of using antibiotics with resistance 0\%}
\label{d7}
\end{table}

\section{Results and discussions}\label{ad}

This study includes a group of 192 dogs affected by otitis externa living in the south of Spain, in which 40 individuals were infected by {\it Pseudomonas aeruginosa}. All samples were aseptically collected and sent to a laboratory to to perform an antibiogram that included 52 different antibiotics.\n

We recall that the goal of this work was to develop an antibiotic panel that could be used as an empirical treatment for dogs infected with {\it Pseudomonas aeruginosa}, a pathogen transmissible to humans, with high multi-resistance capacities and that is considerate the leading cause of nosocomial infections in human medicine. To this end, the well-known SIRI model \eqref{1}-\eqref{3} was used which helped us to provide a priority antibiotic treatment.\n

In this way, the behavior of the different groups of individuals of the population of dogs infected by {\it Pseudomonas aeruginosa} was studied using the SIRI model to observe the dynamics of every group. In addition a code was programmed in {\it Matlab} to obtain the most relevant information from these simulations and then, they were represented in a table where the limit value of susceptible, infected and recovered individuals were included as well as the time at which the corresponding limits were reached.\n

The SIRI model \eqref{1}-\eqref{3} reunited all the necessary characteristics and, after comparing with the possible application of other mathematical models, this is the one that was the most suitable and accurate relating to all the epidemiological data compiled as well as to clinical behavior of the infection. To our knowledge, this is the firs time this specific application of the model is described relating to pathogenic microorganisms dynamics.\n

Especially, it is worth mentioning that this work is quite relevant for mathematical science, since it proves that mathematical modelling should become part of the convenient toolbox of Public Health research and decision-making, specially in the design of models of empirical treatment choices based on epidemiological data.\n

Interestingly, results show that a significant number of antibiotics tested (27,45\%) have null action to {\it Pseudomonas aeuruginosa} infection. In addition, it is remarkable to add that some of them are third generation {\it Cephalosporins} (more specific and expensive, with a more complex development). Moreover, approximately 12\% of the antibiotics included had a resistance rate of almost 100\% and 8\% of them between 50\% and 30\%. Hence, when an empirical antibiotic treatment is provided, the chance of ineffectiveness or partial effect (which is one of the causes that increases the resistance rate) is, at least, 30\%.\n

This fact highlights the potential health problem that presents this multi-resistant microorganisms and the necessity of an accurate tool to prescribe the most specific and effective treatment. On the other hand, almost a third part of the total of antibiotic tested were always effective, however the use of a high number of them are hospital restricted.\n

\begin{figure}[H]
\begin{center}
\includegraphics[scale=0.8]{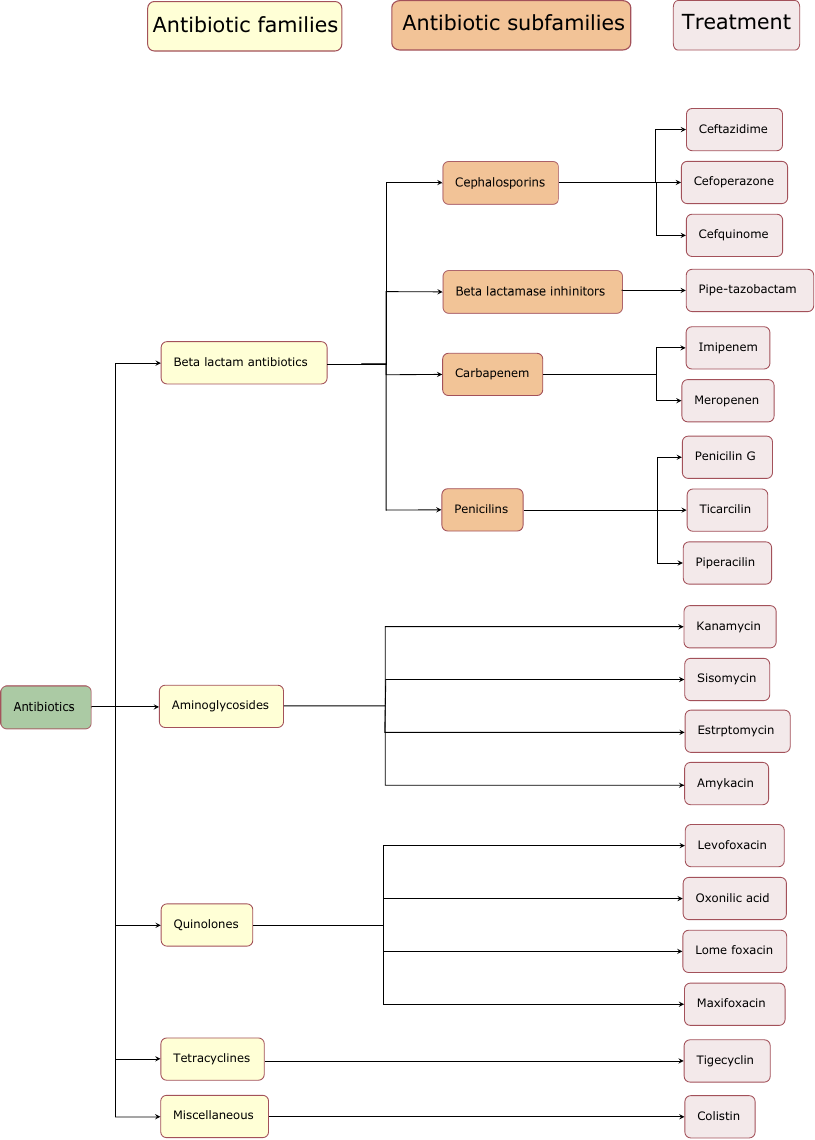}
\end{center}
\mbox{}
\caption{Antibiotic panel recommended to prescribe an empirica treatment against {\it Pseudomonas aeruginosa} infection in Spain}\label{tree}
\end{figure}

Regarding the results obtained, we consider that the SIRI model implemented in the present work can be a very useful tool to predict the dynamic of Pseudomonas aeruginosa infection in dogs, since it provides an antibiotic panel that in case is empirically prescribed, it has an accuracy of efficacy of more than 90\% (Figure \ref{tree}) . Subsequently, it can be used as an efficient model to formulate potential personalized treatment in both human and veterinary medicine, in order not only to save cost, but also to avoid future problems of development of multi-resistance by this bacteria in both humans and animals, because its high level of specificity and efficacy against the infection.\n

Although all the antibiotics included in the recommended panel have an efficacy of more than 90\%, the use of most of them are hospital-restricted, which could present, in this concrete infection and microorganism, a problem in prescribing the treatment in some clinical situations.\n  

The results obtained are in consonance with the epidemiological data and the information that the World Health Organization, as well as numerous Epidemiological Health Services, in human and veterinary medicine, are providing.  Consequently, the present work highlights the severity of the multi-resistance development in pathogens and reinforces the necessity of new tools to implement in the strategies that have to be applied.\n

Figure \ref{tree} provides a selective antibiotic panel in case is necessary to prescribe an efficient empirical treatment to {\it Pseudomonas aeruginosa} infection.

\section{Conclusions}\label{finish}

Multi-resistant infections are an unequivocal public health problem and are responsible for an increasing number of deceased patients in hospitals. In this sense, companion animals can act as a reservoir of this type of microorganisms that can be transmitted to humans.\n

At this point, since no studies of these characteristics have been previously described, it is urgent and critical to develop an efficient tool to minimize the antimicrobial resistance process. Therefore, we have applied an adapted SIRI model as a compatible and potential tool based on {\it Matlab} software. Remarkably, this model allows predicting the behavior of {\it Pseudomonas aeruginosa} infection and is essential to select an effective empirical treatment when culture and antibiogram techniques are no available or the severity of the infection obliges an imminent treatment, providing an antibiotic panel with an accuracy of efficacy of more than 90\%.\n

Our study suggest that this method could be suitable to be used in other infections therapies in other microorganisms and in other countries based on their epidemiological data. Furthermore, it is likely to reinforce the accuracy to provide a more effective treatment of choice, using a model that behaves as a very useful tool to be implemented in Public Health research and decision-making strategies.  

\appendix

\section*{Appendix}

In Appendix we provide the table of antibiotics tested as a treatment.

\begin{center}
\begin{longtable}[H]{|l|l|l|c|}
\hline
\multicolumn{2}{|c|}{Group} & Antibiotic & Resistance (rate) \\ \hline
\endfirsthead
\hline
\multicolumn{2}{|c|}{Group} & Antibiotic & Resistance (rate) \\ \hline
\endhead
\hline
\multicolumn{4}{c}{Continue in the next page}
\endfoot
\caption{Table of all the antibiotics assayed in the antibiograms included in the study}\label{at}
\endlastfoot
Beta lactam & Cephalosporins & Cefalotin & 100\% \\ \cline{3-4}
 antibiotics &  & Ceftazidime & 16,67\% \\ \cline{3-4}
 &  & Cefotaxime & 47,37\% \\ \cline{3-4}
 &  & Cefepime & 21,05\% \\ \cline{3-4}
 &  & Cefoxitin & 100\% \\ \cline{3-4}
 &  & Cefuroxime & 100\% \\ \cline{3-4}
 &  & Cefazolin & 100\% \\ \cline{3-4}
 &  & Ceftazidime/CA & 82,35\% \\ \cline{3-4}
 &  & Cefovecin & 95,24\% \\ \cline{3-4}
 &  & Cefalexin & 100\% \\ \cline{3-4}
 &  & Cefotaxime/CA & 100\% \\ \cline{3-4}
 &  & Ceftiofur & 100\% \\ \cline{3-4}
 &  & Cefpodoxime & 100\% \\ \cline{3-4}
 &  & Cefoperazone & 0\% \\ \cline{3-4}
 &  & Cefquinome & 0\% \\ \cline{2-4}
 & Beta lactamase  & Amoxicilin clav. & 97,50\% \\ \cline{3-4}
 & inhibitors & Ampicilin & 100\% \\ \cline{3-4}
 &  & Piperacillin-tazobactam & 5,88\% \\ \cline{2-4}
 & Carbapenem  & Imipenem & 7,14\% \\ \cline{3-4}
 &  & Ertapenem & 53,85\% \\ \cline{3-4}
 &  & Meropenem & 0\% \\ \cline{2-4}
 & Monobactams & Aztreonam & 20,69\% \\ \cline{2-4}
 & Penicilins  & Piperacillin & 0\% \\ \cline{3-4}
 &  & Ticarcillin & 0\% \\ \cline{3-4}
 &  & Penicillin & 0\% \\ \hline
Aminoglycosides  &  & Tobramycin & 15,15\% \\ \cline{3-4}
 &  & Gentamicin & 23,68\% \\ \cline{3-4}
 &  & Amikacin & 6,25\% \\ \cline{3-4}
 &  & Neomycin & 50\% \\ \cline{3-4}
 &  & Kanamycin & 0\% \\ \cline{3-4}
 &  & Sisomicin & 0\% \\ \cline{3-4}
 &  & Streptomycin & 0\% \\ \hline
Quinolones  &  & Ciprofloxacin & 28,57\% \\ \cline{3-4}
 &  & Enrofloxacin & 34,21\% \\ \cline{3-4}
 &  & Marbofloxacin & 23,81\% \\ \cline{3-4}
 &  & Nalidixic acid & 93,33\% \\ \cline{3-4}
 &  & Pradofloxacin & 28,57\% \\ \cline{3-4}
 &  & Levofloxacin & 0\% \\ \cline{3-4}
 &  & Oxolinic acid & 0\% \\ \cline{3-4}
 &  & Maxifloxacin & 0\% \\ \cline{3-4}
 &  & Lomefloxacin & 0\% \\ \hline
Tetracyclines  &  & Doxycycline & 92,86\% \\ \cline{3-4}
 &  & Tetracycline & 83,33\% \\ \cline{3-4}
 &  & Tigecycline & 0\% \\ \hline
Afenicol &  & Chloramphenicol & 100\% \\ \cline{3-4}
 &  & Florfenicol & 100\% \\ \hline
Miscellaneous  &  & Polymyxin B & 15,79\% \\ \cline{3-4}
 &  & Colistin & 0\% \\ \cline{3-4}
 &  & Fusidic acid & 100\% \\ \cline{3-4}
 &  & Fosfomycin & 100\% \\ \hline
Macrolide &  & Erythromycin & 33,33\% \\ \hline
Sulfamide &  & Sulfonamides+Trimethoprim & 100\% \\ \hline
\end{longtable}
\end{center}

\bibliographystyle{elsarticle-num.bst}

\bibliography{ACLP_submitted}

\begin{thebibliography}{10}
\expandafter\ifx\csname url\endcsname\relax
  \def\url#1{\texttt{#1}}\fi
\expandafter\ifx\csname urlprefix\endcsname\relax\def\urlprefix{URL }\fi
\expandafter\ifx\csname href\endcsname\relax
  \def\href#1#2{#2} \def\path#1{#1}\fi

\bibitem{RB8}
J.~Rubin, R.~Walker, K.~Blickenstaff, S.~Bodeis-Jones, S.~Zhao, Antimicrobial
  resistance and genetic characterization of fluoroquinolone resistance of {\it
  \uppercase{p}seudomonas aeruginosa} isolated from canine infections,
  Veterinary Microbiology 131~(1-2) (2008) 164--172.
\newblock \href {http://dx.doi.org/10.1016/j.vetmic.2008.02.018}
  {\path{doi:10.1016/j.vetmic.2008.02.018}}.

\bibitem{AN6}
V.~Aloush, S.~Navon-Venezia, Y.~Seigman-Igra, S.~Cabili, Y.~Carmeli,
  Multidrug-resistant {\it \uppercase{p}seudomonas aeruginosa}: Risk factors
  and clinical impact, Antimicrobial Agents and Chemotherapy 50~(1) (2006)
  43--48.
\newblock \href {http://dx.doi.org/10.1128/aac.50.1.43-48.2006}
  {\path{doi:10.1128/aac.50.1.43-48.2006}}.

\bibitem{VP16}
D.~van Duin, D.~L. Paterson, Multidrug-resistant bacteria in the community,
  Infectious Disease Clinics of North America 30~(2) (2016) 377--390.
\newblock \href {http://dx.doi.org/10.1016/j.idc.2016.02.004}
  {\path{doi:10.1016/j.idc.2016.02.004}}.

\bibitem{ZM19}
V.~M. Paz-Zarza, S.~Mangwani-Mordani, A.~Mart{\'{\i}}nez-Maldonado,
  D.~{\'{A}}lvarez-Hern{\'{a}}ndez, S.~G. Solano-G{\'{a}}lvez,
  R.~V{\'{a}}zquez-L{\'{o}}pez, {\it \uppercase{P}seudomonas aeruginosa}:
  patogenicidad y resistencia antimicrobiana en la infecci{\'{o}}n urinaria,
  Revista chilena de infectolog{\'{\i}}a 36~(2) (2019) 180--189.
\newblock \href {http://dx.doi.org/10.4067/s0716-10182019000200180}
  {\path{doi:10.4067/s0716-10182019000200180}}.

\bibitem{HC}
J.-E. Hyun, T.-H. Chung, C.-Y. Hwang, Identification of {VIM}-2
  metallo-$\beta$-lactamase-producing {\it\uppercase{p}seudomonas} aeruginosa
  isolated from dogs with pyoderma and otitis in korea, Veterinary Dermatology
  29~(3) (2018) 186--e68.
\newblock \href {http://dx.doi.org/10.1111/vde.12534}
  {\path{doi:10.1111/vde.12534}}.

\bibitem{PR16}
C.~Pomba, M.~Rantala, C.~Greko, K.~E. Baptiste, B.~Catry, E.~{van duijkeren},
  A.~Mateus, M.~A. Moreno, S.~Py\"{o}r\"{a}l\"{a}, M.~Ruzauskas, P.~Sanders,
  C.~Teale, E.~J. Threlfall, Z.~Kunsagi, J.~Torren-Edo, H.~Jukes, K.~Torneke,
  Public health risk of antimicrobial resistance transfer from companion
  animals, Journal of Antimicrobial Chemotherapy (2016) 957--968\href
  {http://dx.doi.org/10.1093/jac/dkw481} {\path{doi:10.1093/jac/dkw481}}.

\bibitem{PY13}
C.~C. Pye, A.~A. Yu, J.~S. Weese, Evaluation of biofilm production by {\it
  \uppercase{p}seudomonas} aeruginosafrom canine ears and the impact of biofilm
  on antimicrobial susceptibility in vitro, Veterinary Dermatology 24~(4)
  (2013) 446--e99.
\newblock \href {http://dx.doi.org/10.1111/vde.12040}
  {\path{doi:10.1111/vde.12040}}.

\bibitem{KM1}
W.~O. Kernack, A.~G. McKendrick, A contribution to the mathematical theory of
  epidemics, Proceedings of the Royal Society of London 115 (1927) 700--721.

\bibitem{KM2}
W.~O. Kernack, A.~G. McKendrick, A contribution to the mathematical theory of
  epidemics, Proceedings of the Royal Society of London 115 (1932) 700--721.

\bibitem{KM3}
W.~O. Kernack, A.~G. McKendrick, A contribution to the mathematical theory of
  epidemics, Proceedings of the Royal Society of London 141 (1933) 94--112.

\bibitem{cc}
T.~Caraballo, R.~Colucci, A comparison between random and stochastic modeling
  for a sir model, Communications on Pure and Applied Analysis 16~(1) (2017)
  151--162.
\newblock \href {http://dx.doi.org/10.3934/cpaa.2017007 .}
  {\path{doi:10.3934/cpaa.2017007 .}}

\bibitem{caraballo-book}
T.~Caraballo, X.~Han, Applied Nonautonomous and Random Dynamical Systems,
  Applied Dynamical Systems, Springer International Publishing, 2016.
\newblock \href {http://dx.doi.org/10.1007/978-3-319-49247-6}
  {\path{doi:10.1007/978-3-319-49247-6}}.

\bibitem{CLSI}
Clinical and Laboratory Standards Institute, \uppercase{P}erformance
  \uppercase{S}tandards for \uppercase{A}ntimicrobial
  \uppercase{S}usceptibility \uppercase{T}esting, 29th Edition (2019).

\end{thebibliography}

\end{document}